\documentclass{svjour3} 
\pdfoutput=1

\usepackage{url}
\usepackage{scrextend}
\usepackage{booktabs} 
\usepackage{diagbox}
\usepackage{adjustbox}
\usepackage{hyperref}
\usepackage{subfigure}
\graphicspath{ {images/} }
\usepackage{graphicx,subfigure}
\usepackage[most]{tcolorbox}
\usepackage{changepage}
\usepackage{dblfloatfix}
\usepackage{multirow}
\usepackage{colortbl}
\usepackage{lscape}
\usepackage{xcolor}
\usepackage{hhline}
\definecolor{dgreen}{rgb}{0.0, 0.5, 0.10}
\definecolor{navyblue}{rgb}{0.0, 0.0, 0.5}

\definecolor{armygreen}{rgb}{0.0, 0.5, 0.13}

\journalname{Empirical Software Engineering}

\newcommand{\surveyquote}[1]{\begin{addmargin}[1em]
{0em}\emph{#1}\end{addmargin}}

\usepackage{pgfplots}
\newenvironment{boxedtext}
    {
    \begin{center}

    \begin{tabular}{|p{0.96\linewidth}|}
    \hline
    }
    { 
    \\ \hline
    \end{tabular} 
    
    \end{center}
       }

\newcommand{\topic}[1]{{\tt #1}}

\newcommand{\ticker} [5]{ \pgfplotsset{ticks=none}
 \resizebox {1.7cm} {.4cm} {
\begin{tikzpicture}
\begin{axis}[
      title=\empty,
      xlabel=\empty,
      ylabel=\empty, axis lines=none,
      yticklabels=\empty,
     bar width=0.7cm , ymin=0, ymax=60,   
    ]
    \addplot[ybar, fill=blue] coordinates {
        (1,{#1}) (2,{#2}) (3,{#3})( 4,{#4})(5,{#5})
    };
    \end{axis}

\end{tikzpicture}
}}

\newcommand{\significant}[1]{ \cellcolor{gray!35} {#1}}

\newcommand{\lessSmall}[1]{\cellcolor{orange!25} \textcolor{red} { \textbf{#1}}}

\newcommand{\less}[1]{\cellcolor{orange!50}\textcolor{red}{\textbf{#1}}}
\newcommand{\more}[1]{\cellcolor{blue!20}\textcolor{dgreen}{\textbf{#1}}}

\newcommand{\moreSmall}[1]{\cellcolor{blue!10}\textcolor{dgreen}{\textbf{#1}}}
\newcommand{\moreLarge}[1]{\cellcolor{blue!30}\textcolor{dgreen}{\textbf{#1}}}

\newcommand{\lessn}[1]{\textcolor{red}{#1}}
\newcommand{\moren}[1]{\textcolor{dgreen}{#1}}

\newcolumntype{C}[1]{>{\centering\let\newline\\\arraybackslash\hspace{0pt}}m{#1}}
\newcolumntype{R}[1]{>{\hfill\let\newline\\\arraybackslash\hspace{0pt}}m{#1}}
\newcommand{\graycell}[1]{\textbf{#1}}

\begin{document}

\title{Understanding the Motivations, Challenges and Needs of Blockchain Software Developers: A Survey}

\author{Amiangshu Bosu \and Anindya Iqbal \and Rifat Shahriyar \and Partha Chakroborty 
}

\institute{A. Bosu \at
              Department of Computer Science, Wayne State University, Detroit, MI, USA \\
              Tel.: +1-313-577-0731\\
              \email{amiangshu.bosu@wayne.edu}           
           \and
           A. Iqbal, R. Shahriar, and P. Chakroborty \at
              Department of Computer Science and Engineering, Bangladesh Uniersity of Engineering and Technology, Dhaka, Bangladesh\\
              \email{anindya@cse.buet.ac.bd, rifat@cse.buet.ac.bd, shuvopartho@gmail.com}
}

\date{Received: November 6, 2018 / Accepted: March 19, 2019}

\titlerunning{Motivations and Challenges of Blockchain Software Developers}
\authorrunning{Bosu et al.}
\maketitle

\begin{abstract}
The blockchain technology has potential applications in various areas such as smart-contracts,  Internet of Things (IoT), land registry, supply chain management, storing medical data, and identity management. Although the Github currently hosts more than six thousand active Blockchain software (BCS) projects, few software engineering research has investigated these projects and its' contributors. Although the number of BCS projects is growing rapidly, the motivations, challenges, and needs of BCS developers remain a puzzle. Therefore, the primary objective of this study is \emph{to understand the motivations, challenges, and needs of BCS developers and analyze the differences between BCS and non-BCS development}.
On this goal, we sent an online survey to 1,604 active BCS developers identified via mining the Github repositories of 145 popular BCS projects. The survey received 156 responses that met our criteria for analysis. 

The results suggest that the majority of the BCS developers are experienced in non-BCS development and are primarily motivated by the ideology of creating a decentralized financial system. Although most of the BCS projects are Open Source Software (OSS) projects by nature, more than 93\% of our respondents found BCS development somewhat different from a non-BCS development as BCS projects have higher emphasis on security and reliability than most of the non-BCS projects. Other differences include:  higher costs of defects, decentralized and hostile environment, technological complexity,  and difficulty in upgrading the software after release. These differences were also the primary sources of challenges to them. Software development tools that are tuned for non-BCS development are inadequate for BCS and the ecosystem needs an array of new or improved tools, such as:  customized IDE for BCS development tasks, debuggers for smart-contracts, testing support, easily deployable simulators, and  BCS domain specific design notations.

\keywords{blockchain \and cryptocurrency \and survey \and bitcoin \and ethereum \and motivation \and challenges}
\end{abstract}

\section{Introduction}
\label{section:introduction}
In 2008, a person or entity under the pseudonym \emph{Satoshi Nakamoto}, published a whitepaper to introduce `Bitcoin,' a  digital currency (aka cryptocurrency) based on an immutable and decentralized public ledger known as \emph{blockchain}~\cite{nakamoto2008bitcoin}. Currently, `Bitcoin' is the leading cryptocurrency with a market cap of over 100 billion USD. On the other hand, the blockchain technology that runs Bitcoin could prove to be much more significant~\cite{swan2015blockchain}, as a blockchain can record transactions between two parties efficiently and in a verifiable and permanent way; thus eliminating the need for the third party in the middle. Moreover, the availability of all the transactions ever completed to all nodes makes a blockchain-based system more transparent than centralized solutions. 
Therefore, apart from its application in cryptocurrency, the blockchain technology has potential applications in various other domains such as smart-contracts~\cite{delmolino2016step},  Internet of Things (IoT)~\cite{huh2017managing}, land registry ~\cite{underwood2016blockchain}, supply chain management~\cite{korpela2017digital}, storing medical data~\cite{azaria2016medrec}, and identity management~\cite{jacobovitz2016blockchain}.

The technological innovation and fundamental changes required in the design, development, and deployment of blockchain have also attracted tremendous interests from the software development community. 
For example, a recent study~\cite{partha2018} from March 2018 reported 3,000 \underline{B}lock\underline{c}hain \underline{s}oftware (BCS) projects hosted on Github\footnote{\url{https://github.com/topics/blockchain}}. In October 2018, within seven months, that number stands at 6,800 (i.e., more than doubled). Unlike traditional development, blockchain developers need to be cautious about malicious actors, secure an immutable and distributed database, and design efficient and reliable protocols withstanding the scarcity of tools and resources which is unavoidable for a new technology. 
The unalterable nature of a blockchain makes the recovery of an error prohibitively difficult or practically next to impossible if the vulnerability is detected after the deployment. Although some other large-scale software such as financial applications requires similar robustness, the rapidly changing blockchain ecosystem adds significant new challenges~\cite{porru2017blockchain}.
The significant differences between traditional software development (i.e., non-BCS) and blockchain oriented development motivated Destefanis \textit{et} al. ~\cite{Destefanis2018BOSE} even to propose a new development paradigm named Blockchain-Oriented Software Engineering (BOSE). 

Although several characteristics of the BCS technology suggest BCS development to be different from a non-BCS development, very few software engineering (SE) research has focused on the former.
Therefore, several questions regarding BCS development still remain a puzzle.
For example, i) \emph{Who are the BCS developers and what are their motivations behind joining BCS development?} ii) \emph{Is BCS development indeed  different from a non-BCS development?} iii) If different, \emph{what are the areas that mark those differences?} iv) Do current development tools and practices that are tuned for non-BCS development satisfy the needs of BCS developers? v) If not, \emph{what are the  tools and techniques that BCS developers are in need?}

Answering these questions would enable the research community and the providers of technical tools to build necessary support to design, develop, test, and deploy BCS applications. Moreover, developers interested to join BCS projects are likely to have valuable guidance for their preparation.
Therefore, this study aims \emph{to understand the motivations, challenges, and needs of BCS developers and analyze the differences between BCS and non-BCS development.}
We also aim to conduct a comparative analysis of BCS with non-BCS development to pinpoint specialized tasks that may benefit the former.

On these goals, we designed and sent an online survey to 1,604 BCS developers gathered via mining the Github repositories of 145 BCS projects. 
 A survey is an ideal instrument for this study as current BCS developers have first-hand experiences of their challenges and needs. 
The survey received 156 responses from BCS developers that met our criteria for analysis. 
We adopted a systematic qualitative analysis approach to build a coding scheme for the open-ended responses. Using a qualitative analysis software, multiple coders independently assigned codes to each response and achieved a `substantial' inter-rater reliability ( Cohen's $\kappa =0.62$)~\cite{Cohen1960}. 
We reported partial results of the survey in a recent publication~\cite{partha2018}, which explored only the software development practices (i.e., verification and validation, task assignment, requirement analysis, and communication and collaboration) of BCS projects. On the other hand, this publication focuses on a different set of questions none of which was included in our prior article.

The primary contributions of this study include:
\begin{itemize}
\item A better understanding of  BCS developers' motivations, challenges, and needs;
\item A comparative analysis of BCS with non-BCS development; 
\item A comparative analysis of the results of our survey with two prior SE surveys to set our results into the perspective of the software development realm.
\item Potential directions for SE researchers to build supports for BCS development; and
\item A characterization of the contemporary BCS development community.
\end{itemize}

The remainder of the paper is organized as follows. 
Section~\ref{sec-background} provides a background on blockchain and a overview of a prior motivating study.
Section~\ref{sec-RQ} introduces the research questions of this study. 
Section~\ref{sec-research-methodology} describes our research methodology. 
Section~\ref{sec-demographics} describes the demographics of our respondents.
Section~\ref{sec-results} presents the results of this study. 
Section~\ref{sec-discussion} discusses the implications of our findings. 
Section~\ref{section-threats} describes the threats to the validity of our results. 
Finally, Section~\ref{sec-conclusion} concludes the paper.

\section{Background}
\label{sec-background}
This section provides a brief overview of the key concepts on blockchain and a prior SE study that assisted in designing our survey questions to compare BCS development against non-BCS.

\subsection{Blockchain}

    \begin{figure}
 \centering    \includegraphics[width=\columnwidth]{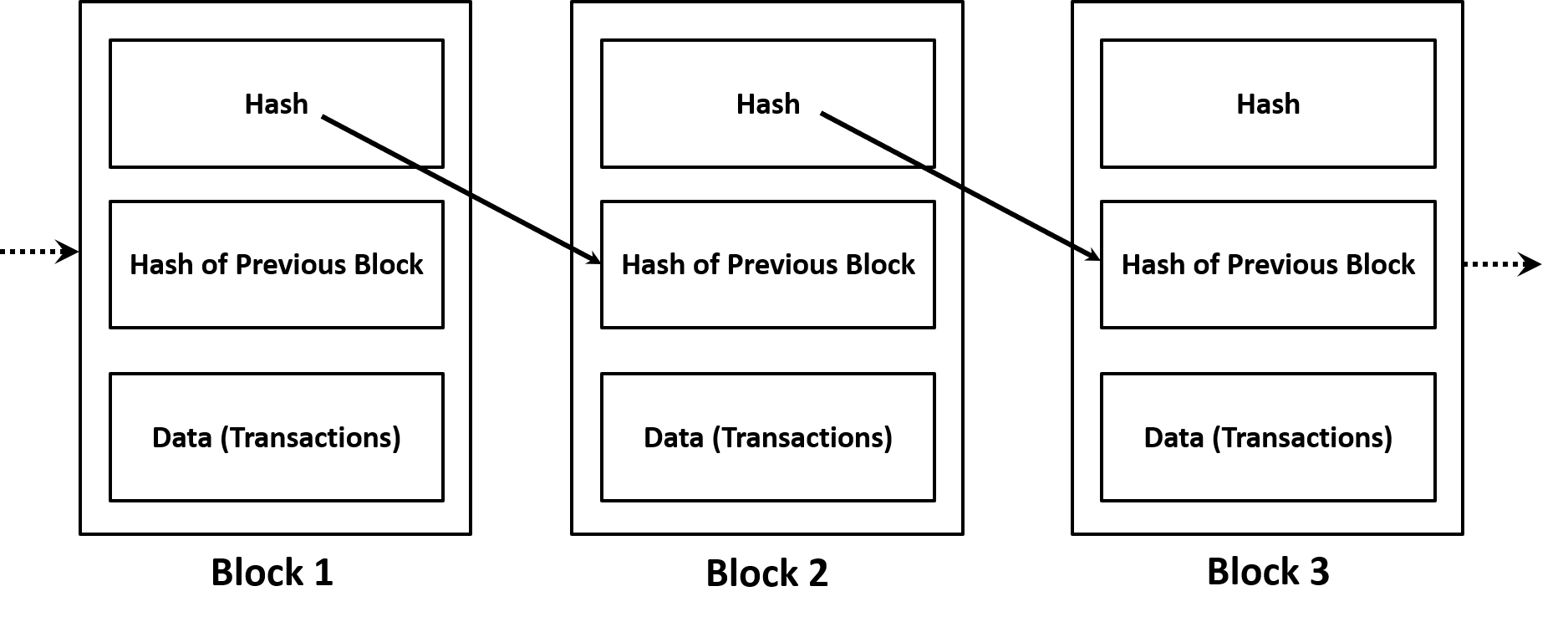}
    \vspace{-16pt}
	\caption{Simplified diagram of a blockchain}
	\label{fig-blockhain}
    \end{figure}
 
 \topic{Blockchain} is a decentralized, peer-to-peer, transparent, immutable, and append-only data storage. It keeps a permanent record of writes called \topic{transactions}.
 Multiple transactions are grouped in \topic{blocks}. 
Each block in a blockchain contains its hash computed using a well-known hashing or proof-of-work~\cite{jakobsson1999proofs} algorithm (e.g., SHA256, ethash, and equihash) and the hash of the previous block called parent block (Figure~\ref{fig-blockhain}). The first block in a chain is called the  \topic{genesis block}, which does not have any parent. Each block's hash is calculated based on its data, current timestamp and the hash of its parent block. Any change in a block's data causes alteration of its hash and invalidates all the subsequent blocks, and the tampering becomes immediately evident to every member node of the chain.  Hence, to compromise a blockchain, collusion of the majority of the network is required which is impractical in case of a large blockchain~\cite{narayanan2016bitcoin}. Therefore, blockchain is a chain of blocks where the blocks are irreversible and immutable.    

All the nodes in a blockchain network participate simultaneously in finding the next block to write. This process is called \topic{mining}, where the nodes calculate a hash value by adding a nonce (i.e., a random value) to a list of transactions waiting to be added to the blockchain. To be eligible as the next block, the hash must be smaller than an agreed-upon value (known as \topic{difficulty}), and the nodes continue calculations using different nonces until they find a nonce that generates a hash satisfying the `difficulty.' The node finding a new block will broadcast it to all other nodes in the network to confirm the correctness of this new block. Once confirmed, the new block is added to the blockchain, and each of the transactions contained in the block is considered \topic{verified}~\cite{chuen2015handbook}. The finder is usually rewarded with a pre-defined number of tokens, known as \topic{block reward}.  The difficulty of the next block is determined by the network with a pre-defined algorithm.

There is no central control over the operation of a blockchain. The underlying philosophy is that no single participant or group of participants can control the infrastructure and all the participants in the network have an equal role to play. In the absence of a central controller, the transactions are mediated by the member nodes using a \topic{consensus} protocol, which ensures that all the nodes have an identical copy of the blockchain. A new block is considered verified only after the majority of the member nodes vote it as true and trustworthy using the consensus protocol. A blockchain's security is based on the assumption that tampering would have to happen across the majority of the nodes (aka \topic{51\% attack}) of a network in the same way simultaneously. So once a blockchain network achieves critical mass,  altering a blockchain posthoc becomes infeasible.

In the context of the blockchain, public key cryptography~\cite{garfinkel1996public} ensures the integrity and authenticity of any message/transaction. Each node owns a pair of asymmetric encryption keys~\cite{simmons1979symmetric}, where the \topic{public key} is broadcast to all relevant nodes but the \topic{private key} is kept secret. A sender signs messages with its own private key and a receiver verifies the integrity of the message by decrypting it with the sender's public key. In cryptocurrency applications, the public key of a user also acts as his/her \topic{account address}. Therefore, a user must sign outgoing transactions using his/her private key. A miner node would verify an outgoing transaction from an account only when it can authenticate the transaction using the owner's public key.

One of the recent innovative applications of blockchain is \topic{Smart-contracts}, which are self-executing contracts with the terms of the agreement between buyer(s) and seller(s) of transactions written using lines of code instead of a legal language. Smart-contracts permit trusted transactions and agreements to be carried out among different anonymous parties without the need for a central authority, legal system, or external enforcement mechanism. Since a smart-contract, once deployed, lives on a distributed and decentralized blockchain network,  it remains traceable, transparent, and irreversible. 

\subsection{Game Development vs. Traditional Development at Microsoft}
\label{ms-study}
One of the objective of this study is to identify the differences between BCS and non-BCS development. The design of our survey questions to investigate this objective is motivated by a prior study at Microsoft (referred as `MS study' hereinafter) by Emerson \textit{et} al.~\cite{murphy2014cowboys}. The primary objective of the MS study was to compare game development against traditional software development. On this goal, the MS study interviewed 14 developers who had both game and non-game development experiences. Those interviews focused on topics from the 10 areas in the Software
Engineering Body of Knowledge (SWEBOK) as well as general work
features from applied psychology~\cite{humphrey2007integrating}. After analyzing the interview transcripts, the authors selected 28 statements to assess the differences between game and non-game development at Microsoft. 

In a survey of Microsoft developers from three different domains (i.e., Game, Office, and Other), each of the respondents were asked to rate each of the selected 28 statements on a 5-point Likert scale from `Strongly Disagree' to `Strongly Agree'. The results of the survey identified differences between game and non-game development in terms of i) having clear requirements, ii) using agile development methods, iii) valuing creativity, iv)  communicating with non-engineers, v) team compositions, and vi) taking pride for developed software. The empirically developed 28 statements spanning various software engineering as well as general work feature is another key contribution from this study, since these statements can be reused to compare the opinions of software developers from two different domains. 
\section{Research Questions}
\label{sec-RQ}
The primary objective of this study is \emph{to understand the motivations, challenges, and needs of BCS developers and analyze the differences between BCS and non-BCS development.} We aim to achieve this goal based on four specific research questions. We also explore one additional research question to characterize the contemporary BCS development community. Following subsections introduce the five research questions with a brief motivation behind each question. 

\subsection{Personal Characteristics of the BCS Developers}
Characterizing the Open Source Software (OSS) developers has drawn interests from the researchers~\cite{DAVID2008364,linus2005,hars2001working,lakhani2003hackers,Mockus2002} to understand the distribution of expertise, the strength (and durations) of their attachments to particular projects, and the recruitment and retention of newcomers. However, such a characterization for BCS developers is currently missing. 
There is a common concept that blockchain development community is dominated by libertarian and anarchist groups who consider it as a means of removing control from an imposing authority ~\cite{ideologyBlog}. The contributors like to create a system that regulates itself and provide advantages to those willing to take part in it~\cite{Reigers2018ideology}. Hence, it is worth formal study the demographics of the BCS developers regarding their age, gender, education, and general software development experience and see if it differs from non-BCS. Also, the future participants of BCS development are likely to have idea about the characteristics of existing community who are involved in the industry with large stake, wide variety of motivation and ethical consideration. Since this understanding has importance, our first research question is:

\vspace{4pt}\textbf{RQ1:} \emph{Who is contributing to BCS projects?}

\subsection{Motivations of BCS Developers}
Much research has focused on understanding the motivations of 
OSS developers~\cite{hars2001working,hertel2003motivation,Lee2017,roberts2006understanding,lakhani2003hackers} and have identified five primary categories motives as following:
\begin{itemize}
    \item \emph{Intrinsic motivations} refer to a person's desire to feel competent and self-determined. Intrinsic motives are directly linked to the emotions of interest and enjoyment~\cite{deci1985intrinsic}. Examples of such motives include fun, hobby, self-interest, and feeling competent.
    \item \emph{External rewards} refer to direct or indirect incentives, which include   monetary compensation, benefits through software usage,  or prospects for career growth~\cite{lakhani2003hackers}.
    \item \emph{Ideology} includes the norms, beliefs, and values shared among the developers of an OSS project~\cite{stewart2006impact}. For example, the Free Software Foundation was established on the ideology of providing users with freedom to use, modify, and redistribute software~\cite{stallman2003free}.
    \item \emph{Community recognition} corresponds to a person's needs for belonging and love.  A developer desiring to identify him/herself as a member of an OSS community seeks that recognition from other community members~\cite{hars2001working}.
    \item \emph{Learning} opportunities offered by OSS projects help  increasing one's `human capital' (i.e., personal skills, capabilities, and knowledge) by means of education, training, learning, and practicing. Since these human capital gains eventually leads to better job opportunities, higher  salaries, and more fulfilling jobs, learning is one of the motives among  many OSS developers, especially newcomers~\cite{hars2001working}.
\end{itemize}
The results of those studies found ``intrinsic motivations'' ranking top among OSS developers. Although most of the BCS projects are also OSS projects, large inflows of cash through ICO (Initial Coin Offerings) or token sales~\cite{ADHAMI2018}, which are very common for BCS projects are rare for non-BCS OSS projects. Many of the early BCS developers / investors have garnered significant financial rewards from the recent boom of the cryptocurrency market.
Therefore,  it won't be surprising if external rewards are the primary motives for many of the BCS developers. Our next research question tries to find out whether BCS developers' motivations are similar to non-BCS OSS developers or they are attracted by potential financial gains. 
Understanding the motivations of BCS developers is important since it will help to identify prospective joiners, which may form synergies with a BCS community. Hence, our next research question is:

\vspace{4pt}
\textbf{RQ2:} \emph{What are the primary motivations of BCS developers?}

\subsection{Differences Between BCS and non-BCS development}
Although BCS projects possess many traits of traditional OSS projects, we expect some differences between BCS and non-BCS development due to several characteristics (e.g., the immutability of data, hostile environment, and difficulty of upgrading the software after deployment) that distinguish the BCS domain from the non-BCS one. Since identifications of the differences between BCS and non-BCS development will allow assessing the applicability of various traditional SE tools and techniques for BCS projects, we seek to find out:

\vspace{4pt}
\textbf{RQ3:} \emph{What are the differences between BCS and  non-BCS development?}

\subsection{Challenges of BCS Development}
Most of the innovative applications of the blockchain technology such as smart-contracts and distributed applications (aka dapps) are at nascent stages. The BCS development landscape is also changing at a rapid pace with projects fiercely competing with each other to emerge as the market leader~\cite{Krafft-2018}. Although the high costs of bugs mandate high reliability from a BCS, developers face tremendous pressures from the investors to release the product. These scenarios coupled with the unique characteristics of the BCS domain pose several challenges to the BCS developers. We believe identifications of the primary challenges of BCS development will provide prospective joiners with guidance for their preparation and encourage research to mitigate those challenges. Hence our next research question is:

\vspace{4pt}
\textbf{RQ4:} \emph{What are the primary challenges of BCS development?}

\subsection{Tools that BCS Developers Need}

 The difficulty of testing BCS ranks among the top of challenges encountered by BCS developers~\cite{porru2017blockchain,testingbcs}. Current testing tools cannot simulate testbeds to simulate the distributed and hostile execution environment of a BCS. Moreover, smart-contract development, which is gaining popularity lacks supporting tools~\cite{clack2016smart}. The requirement of tools, once clearly understood, will lead to the development of supporting tools to design, develop, test, and deploy BCS applications. Since contemporary BCS developers have the first-hand knowledge of their needs for supporting tools, we inquire:

\vspace{4pt}\textbf{RQ5:} \emph{What are the tools that BCS developers currently need?}

\section{Research Methodology}
\label{sec-research-methodology}
Since the five research questions of this study are geared towards gathering the opinions of BCS developers, we chose a survey as our research instrument. 
The remainder of this section describes the survey design, our participant selection criteria, pilot testing, data collection, and qualitative data analysis.

\subsection{Survey}
\label{sec_survey}
Our goal in designing the survey was to keep it as short as possible, while still gathering all of the relevant information.
For the current paper, we only consider a subset of the survey questions, while question focusing on software development practices of BCS projects were reported in a recent publication~\cite{partha2018}.
Table~\ref{table:survey_questions} lists each survey question included in this paper, the research question that motivated its inclusion, and the answer choices provided.
Questions indicated with a `D,' rather than a `RQ\#' were included to gather demographics about the respondents.
For the questions that were open-ended, there are no specified answer choices.
Several of the open-ended and demographics questions were inspired by previous surveys~\cite{murphy2014cowboys,hertel2003motivation,von2012carrots,hars2001working,lakhani2003hackers,west2006challenges}.

To compare BCS with non-BCS development, the survey asked   the respondents (Q13 in Table~\ref{table:survey_questions}) to rate their agreements for 16 statements  on a  5-point Likert scale from `Strongly disagree'  to `Strongly agree.' These statements (`Statement' column in Table~\ref{table:comparison}) were adopted from a prior SE survey at Microsoft  (Section~\ref{ms-study}) to assess the differences between the game and non-game development~\cite{murphy2014cowboys}. While the MS study had 28 statements, our survey did not include the statements regarding the company or manager as BCS projects are mostly community driven. 

\begin{table*}
	\centering
    \caption{Survey Questions}
	\label{table:survey_questions}
\resizebox{\linewidth}{!} {
\begin{tabular}{p{.02\textwidth}p{.02\textwidth}p{.55\textwidth}p{.35\textwidth}}
\hline
\textbf{\#} & \textbf{RQ*} & \textbf{Question Text} & \textbf{Answer Choices} \\
\hline
Q1 & RQ1 & How old are you? & [\#]  \\
Q2 & RQ1 & Which gender do you identify yourself with? & [Male, Female, Prefer not to disclose] \\ 
Q3 & RQ1 & What is your highest level of education? & [High school, Bachelors, Masters, Ph.D.] \\
Q4 & RQ1 & How many years of software development experiences do you have? & [Less than a year, between one to five years, between six to ten years, more than ten years] \\
Q5 & D & How many years have you been developing blockchain software?& [Less than a year, between one to two years, between three to five years, more than five years] \\
Q6 & D & What is your primary Blockchain software project (i.e. the project that you have spent most of your time) & [\#]  \\
&\\
Q7 & D & Do you receive any direct compensation (i.e, salary or cryptocurrency) from your primary project? & [I get directly paid with a FIAT currency, I get compensated with shares, tokens, or cryptocurrency, I receive both FIAT salary and tokens/shares, No, I do not receive any direct compensation] \\
Q8 & D & Approximately, how many pull requests have you submitted to your primary project? & [Less than 10,  Between 11 to 30, More than 30] \\
Q9 & D & Approximately, how many hours on average do you spend per week on your primary project? & [Less than 5, between 6 to 10, Between 11 to 20, Between 21 to 35, I work full time] \\
Q10 & RQ2 & What are your motivations to contribute to your primary project?  &\\
Q11 & RQ3 &  Based on your experiences, what are the primary differences between blockchain and non-blockchain software development ? & [\#]  \\
Q12 & RQ3 & \textit{ The following statements aim to compare blockchain and non-blockchain software development. For each of the following items please rate how you agree or disagree with
that statement}.& [Strongly disagree, Disagree, Neutral, Agree, Strongly Agree]  \\
&  &Total 16 statements ( shown in Table~\ref{table:comparison}) &  \\
Q13 & RQ4 & What are the most challenging aspects of blockchain software development? &\\
Q14 & RQ5 &  Please describe the type of tools that you currently do not have, but if implemented, can greatly help your blockchain software development activities. & [\#]  \\
\hline
\multicolumn{4}{l}{\small{\textit{* `RQ\#' numbers refer to the research question that motivated the inclusion of the survey question}}}\\
\multicolumn{4}{l}{\small{\textit{* Questions indicated with a `D', were included to gather demographics about the respondents. }}}\\

\end{tabular}
}
\end{table*}

\subsection{Participant Selection}
To ensure valid results, we only surveyed BCS developers with sufficient experience. We  identified 145 BCS projects based on following four criteria:
\begin{itemize}
\item Tagged under at least one of the following six `topics'\footnote{\url{https://blog.github.com/2017-01-31-introducing-topics/}}: \topic{blockchain}, \topic{cryptocurrency}, \topic{altcoin}, \topic{ethereum},     \topic{bitcoin},  and \topic{smart-contracts}.
\item `Starred' by at least ten users.
\item Have at least five distinct contributors.
\item A manual verification of the repository confirmed it as a BCS project.
\end{itemize}

 We used Github API\footnote{\url{https://developer.github.com/v3/}} to identify 1,604 contributors, each of whom had submitted at least five changes to one of those 145 projects. 
 We mine the Git commit logs of the identified 145 projects to gather the email addresses of those 1,604 active contributors. We also got the survey questions, consent form, participant selection strategy, solicitation email, and data management reviewed and approved by our university's  Institutional Review Board (IRB).

\subsection{Pilot Survey}
To help ensure the understandability of the survey, we asked Computer Science professors and graduate students with experience in SE and experience in survey design to review the survey to ensure the questions were clear and complete.
The feedback only suggested minor edits.
The changes we made include: adding more answer choices to several questions and adding clarifying examples to three questions.

\subsection{Data Collection}
 On December 13, 2017, we sent each of the 1,604 BCS developers in our list a personalized email mentioning the BCS repository that we mined to obtain his/her email address with a link to the survey hosted on Qualtrics~\cite{snow2013qualtrics}. We also asked the respondents through both the solicitation emails and a reminder in the survey to answer our questions based on his/her personal experiences with the BCS project where we obtained his/her email address.
Since 62 of our solicitation emails bounced, we were left with at most 1,542 potential participants, assuming all other emails actually reached their intended recipient. On December 21, 2017, we sent a reminder email. We closed the survey on January 5, 2018; after the response rate slowed to almost no response each day.

Data from the survey link created with Google's URL shortener showed a total 358 clicks on the survey URL ($\approx$23\% of the invitations). Out of those clicks, 200 people took the survey with a  response rate of $\approx$13\% (200/1542). 
As most of the questions were optional, many respondents skipped some of the questions.
Only 115 respondents answered all the questions. 
After the exclusion of the 44 responses that did not answer either at least 75\% of the questions or at least one open-ended question, we were left with 156 responses for analysis.

We also collect the developer demographics data from the \emph{2018 Stack Overflow Annual Developer Survey} (referred as the `SO Survey' hereinafter)~\cite{so-survey}. 
StackOverflow has been running the annual developer survey since 2011. The primary objective of the SO Surveys are to learn who contemporary developers are and what they need. These surveys cover a wide range of developer demographics such as age, education, location, gender, role, experience, ethnicity, and favorite technologies. Since the `2018 SO survey' was responded by total 98,855 developers  from 183 countries worldwide, it is an accurate overview of the active software developers worldwide. Therefore, a comparison against the demographics of the SO Survey will enable us to identify if the BCS community is underrepresented or overrepresented by certain groups. 

After contacting the authors of the MS survey, we were able to obtain their dataset, which allows us to compare and contrast BCS development with three non-BCS domains (i.e., \emph{Games}, \emph{Office}, and \emph{Other}) from Microsoft (MS).
\subsection{Qualitative Analysis Process}
\label{sec:Open-Ended-Questions}
For the open-ended questions, we followed a systematic qualitative data analysis process. First, two of the authors independently extracted the general themes from the first 75 responses to each question. Using those themes, the authors had discussion
sessions to develop an agreed-upon coding scheme for each question. Using this coding scheme, another author went through the remaining answers to determine any additional codes that need to be added.

With this scheme, two of the authors independently coded each response using the Coding Analysis Toolkit (CAT)~\cite{lu2008rigor} software. The coders could also add new codes, if necessary.  We computed the level of inter-rater reliability of the manual coding process using Cohen's kappa~\cite{Cohen1960}, which was measured as 0.62.
While there is no universally accepted `good' kappa, values between 0.61 to 0.80 are generally recognized as `substantial agreements'~\cite{Landis-Koch:1977}. 
We used CAT to identify the discrepancies in coding and had discussion sessions to resolve all conflicts.
Once we completed the coding process, we transferred the data into SPSS~\cite{spss} for further analysis along with the quantitative data.

As a result of the coding process, a large number of codes emerged from each of the open-ended questions. To help with our analysis,
we had discussion sessions to identify the codes that express similar themes.   We grouped those codes into a smaller number of high-level categories with each category including one or more codes expressing a similar theme. Table~\ref{table:codes} in the appendix shows the codes that emerged from our open-coding of the four survey questions and the categories that we assign each code to for the four open-ended questions included in this paper.

\section{Demographics}
\label{sec-demographics}

To provide a proper context for the results, this section describes the demographics of the projects represented by the respondents and of the respondents themselves.

\subsection{Projects Represented}
Table~\ref{table:projects} provides the results to Q6 (Table~\ref{table:survey_questions}) about respondents' primary projects. The number in parenthesis represents the number of respondents who listed that project. Our respondents represent 61 different BCS projects. The Coin Development Index~\cite{coin-dev-index}, which tracks the top BCS projects, indicates our respondents representing 18 out of the top 25 projects. Also, 37\% of our respondents coming from the top ten projects indicates the participation of the top BCS developers in our survey.

\begin{table*}
	\centering
    \caption{Projects represented by our respondents. The numbers in parentheses under the columns `Multiple occurrences' represent the number of respondents who listed that project. Only one respondent from each of the projects listed under `Single occurrences'.}
	\label{table:projects}
	\resizebox{\linewidth}{!} {
\begin{tabular}{|l|l|l|l|l|}
\hline 
\multicolumn{2}{|c|} { \graycell{Multiple occurrences}} & \multicolumn{3}{c|}  { \graycell{ Single occurrences}} \\ 
\hline 
Ethereum (22) & Bitcoin (9)  & Ambisafe & Basic Identity Token & Bytom 
\\ \hline
 Bitshares (7) & Monero (6) & Cpuminer&  Dash & Distense  
\\ \hline
Sia (6) & Waves (5) &  DNSChain & Ebets & ESKU 
\\ \hline
 Solidity (5) & Lbry (4) & Etherbet &Etherplay & Fabric Labs 
\\ \hline
 Ripple (4) & Nem (3) & Golem & Haskoin & ZeroLink  
\\ \hline
 Cardano (3) & Decred (3) & Icofunding & Ind & Iroha  
\\ \hline
 EOS (3) & Hyperledger (3) & JS Miner & Keyrun & Libsnark  
\\ \hline
 IOTA (3) &Factom (2)  & LiteCoin & Payroll System & PHP-Mpos   
\\ \hline
 Feather coin (2) & Lisk (2) & Populus  & Progmathon & Pycoin 
 
 \\ \hline 
Metamask (2) & Namecoin (2) & Shapeshift &  Snapcoin & Status  
\\ \hline

 Neo (2) & Remix IDE (2) & Steller & Storj  & Swifty 
\\ \hline

 Stratis (2) & Trezor (2) &Vandal  & Vcash & Viper
 \\ \hline
 Zcash (2)&  \multicolumn{4}{l|}{ Undisclosed /Private (14)}
\\ \hline
\end{tabular} 

}
\end{table*}

\subsection{Respondents' Demographics}
\label{dem-resp}

\begin{figure*}
\centering     
\subfigure[BCS development experience]
{\label{fig:bcs-experience}
\includegraphics[width=0.34\linewidth]{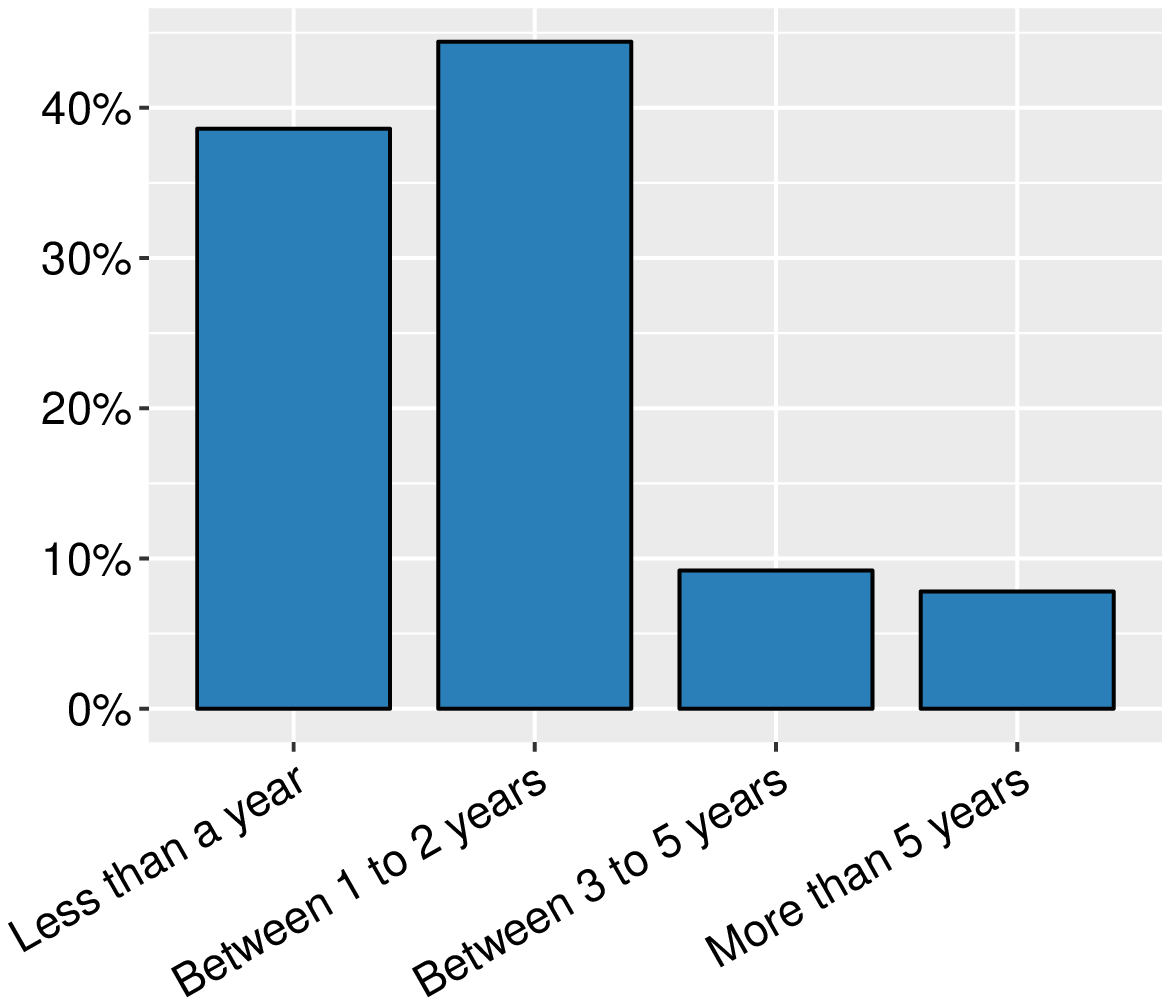}}
~
\subfigure[Type of financial compensation]
{\label{fig:compensation-type}
\includegraphics[width=0.34\linewidth]{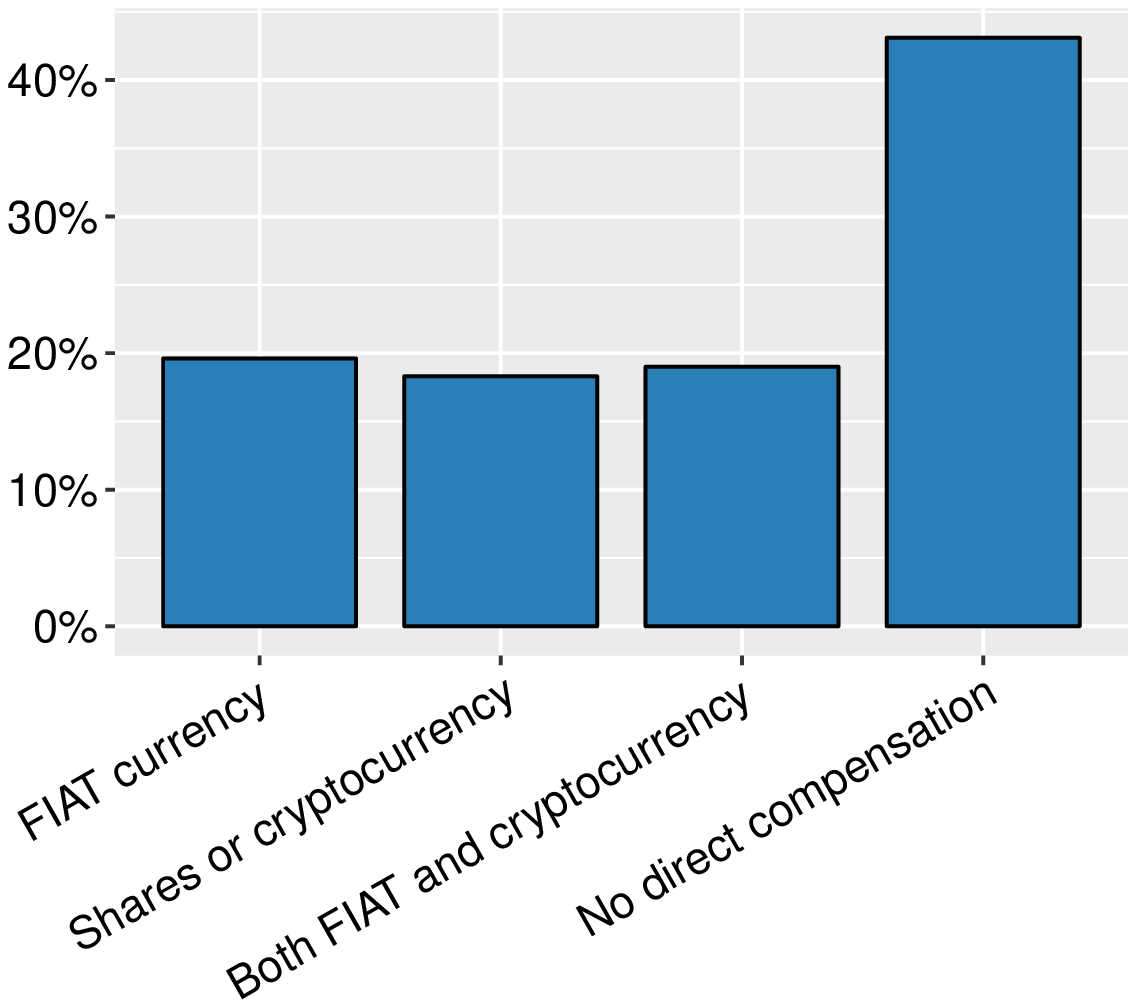}}
~

\subfigure[Total number of code commits in BCS projects]
{\label{fig:pull-requests}
\includegraphics[width=0.34\linewidth]{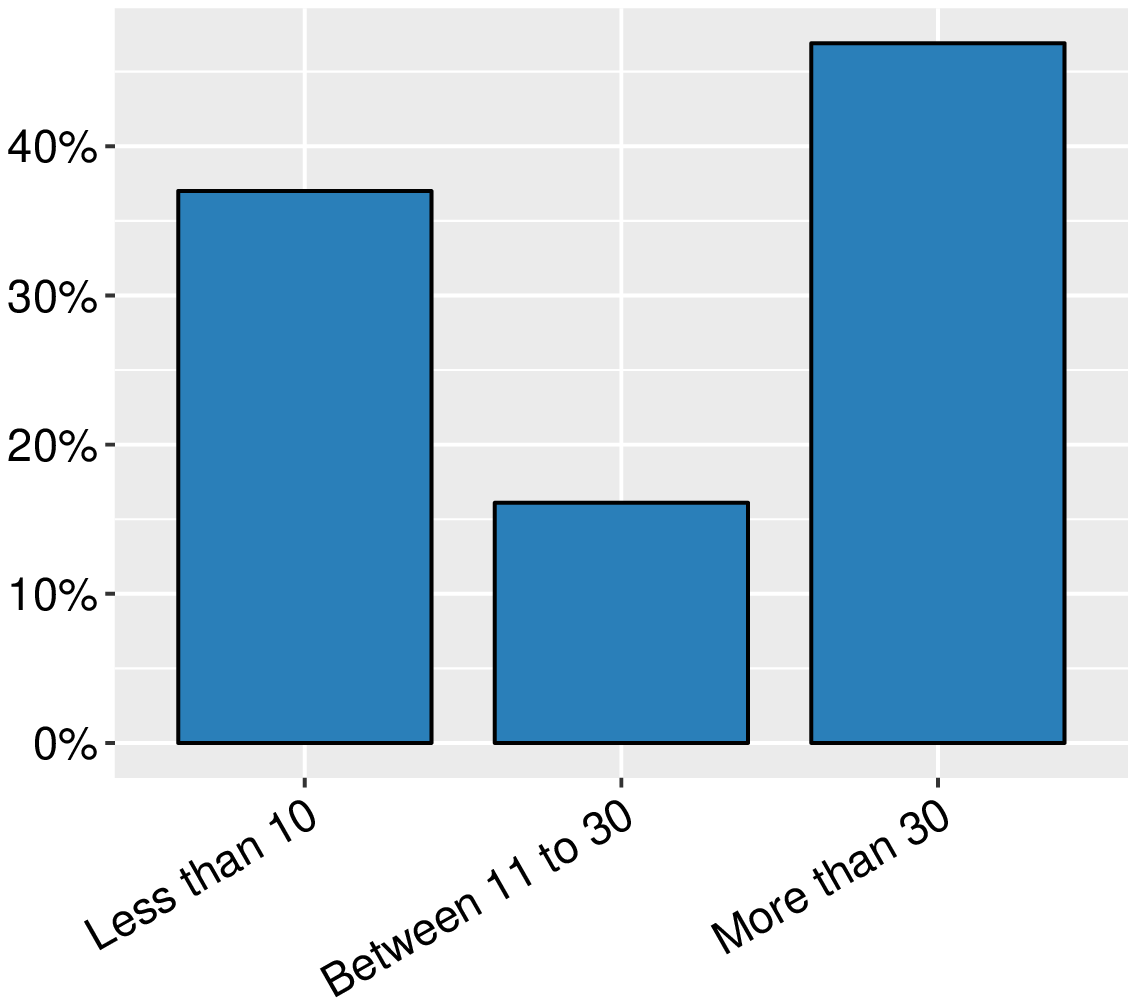}}
~
\subfigure[ Average number of hours per week spent on BCS development]
{\label{fig:num-hours}
\includegraphics[width=0.34\linewidth]{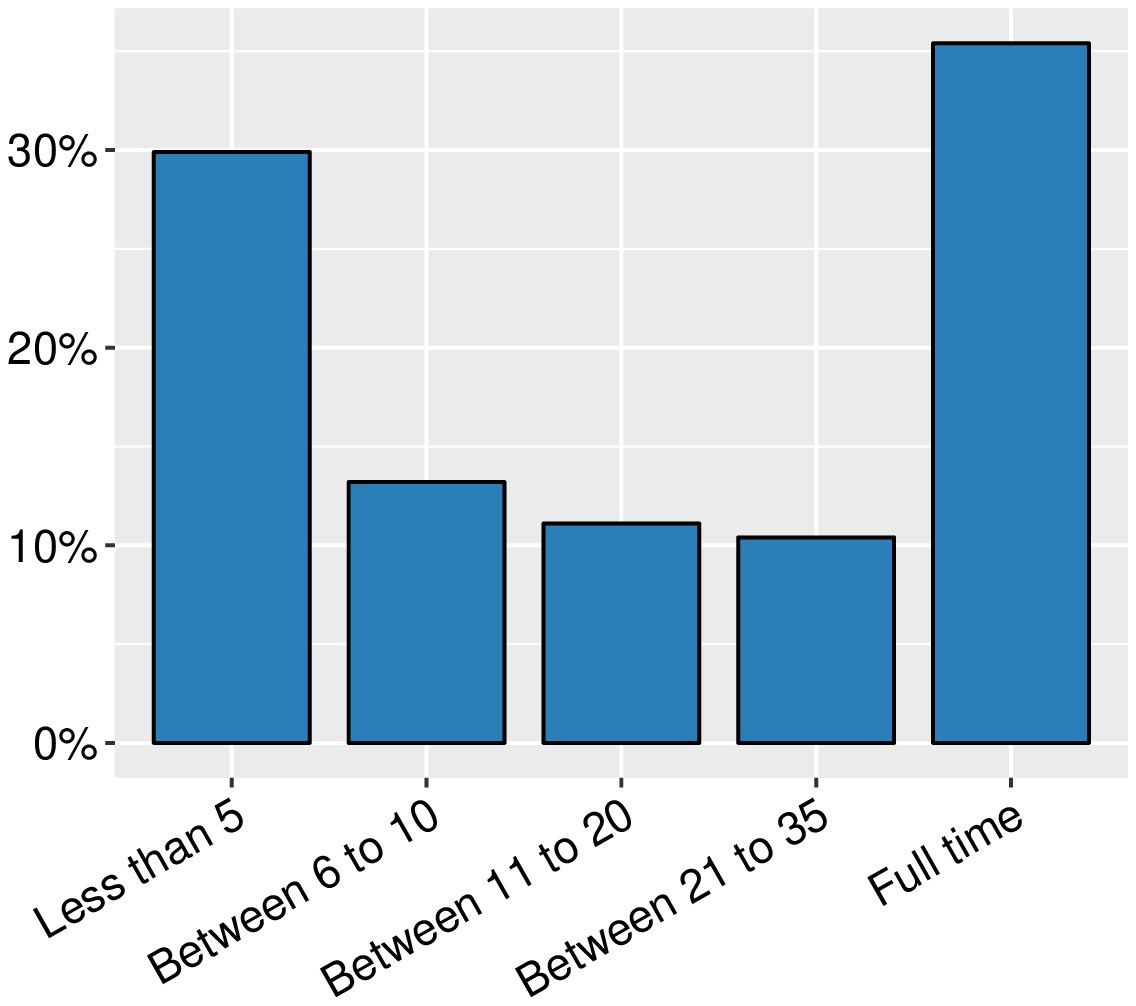}}
\caption{Demographics of the respondents}
\label{fig:dev-interest}
\end{figure*}

In response to Q5, 81.4\% of our respondents indicated having less than 2 years of BCS development experiences, while 37.8\% had less than a year (Figure~\ref{fig:bcs-experience}). 
In terms of compensation (Figure~\ref{fig:compensation-type}), 43.1\%  of our respondents do not receive any direct financial benefits. The ratio of volunteers among our respondents is similar to the ratios reported in prior studies~\cite{bosu2017process,Bosu-etal:JSS}. Among the respondents receiving direct financial compensations, 37.3\% reported receiving shares, tokens or cryptocurrencies that may further motivate them to spend efforts to make their projects successful.

In terms of the number of contributions to a BCS project (Figure~\ref{fig:pull-requests})  57.6\% of our respondents have made more than 10, while 42.9\% had submitted more than 30. On the other hand, 42.7\% of our respondents spend at least 20 hours a week on a BCS project, and 32.7\%  were working full time (Figure~\ref{fig:num-hours}). Combining our respondents' number of commits and number of hours per week spent in BCS projects, we conclude that our respondents include a sample of active BCS developers from the top BCS projects who are qualified to provide valuable insights for the goals of this study.

\section{Results}
\label{sec-results}

The following subsections describe the results of our survey by answering  the five research questions introduced in Section~\ref{sec-RQ}.
To help clarify the results, we also include excerpts from the qualitative responses to the open-ended questions.
Each of the excerpts is followed by a number representing a unique identifier for the respondent who expressed that opinion. For example, [\#5] indicates a response from  respondent number 5. 
In a qualitative analysis, each open-ended response could match multiple codes.
Therefore, the sum of the percentages can be greater than 100\%. We conclude each of the subsections with a brief discussion of the key takeaways from the results of each research question.


\subsection{RQ1: Who is contributing to BCS?}
\label{sec-RQ1}

    \begin{figure}
 \centering    \includegraphics[width=\columnwidth]{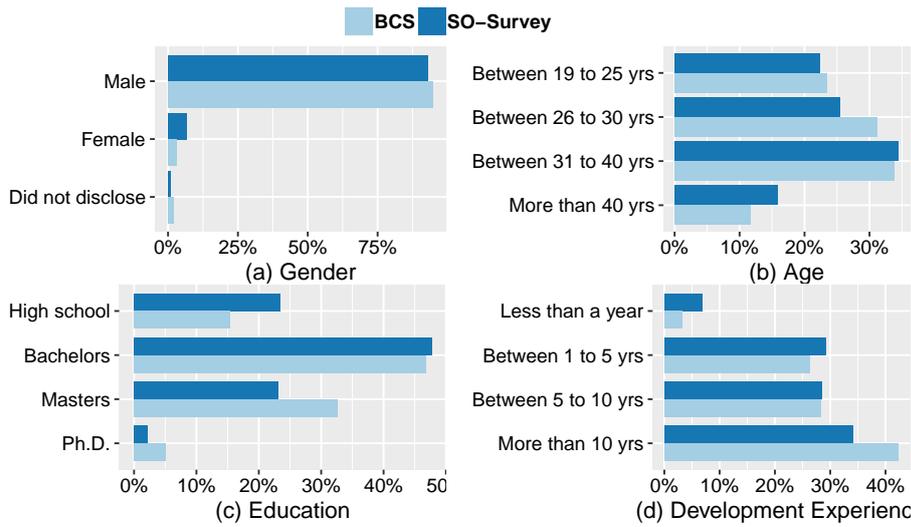}
    	\caption{Personal characteristics of the BCS developers}
	\label{fig-RQ1}
	   \end{figure}

Figure~\ref{fig-RQ1} shows the personal characteristics of the BCS developers in terms of their  age (Q1), gender (Q2),  education (Q3), and software development experience (Q5).
We also compare these characteristics with the results from the  that includes the demographics of 98,855 developers around the globe.

Around 95\% of our respondents are males compared to only 3\% females (Figure~\ref{fig-RQ1}(a)). These numbers suggest that the ratio of females in BCS development may be lower compared to the the software development community as reported in the SO survey (6.7\%). 

In terms age (Figure~\ref{fig-RQ1}(b)), our respondents are younger compared to the general software developer population. The distributions are noticeably different for two age groups. First, developers, who are  aged between  26 to 30  years represent 31\% of the BCS developers, compared to 25.4\% general software developer population coming from the same age group. On the other hand, although 15.9\%  general software developer population are aged 40 years and higher, only 11.7\%  of the BCS developers belong to that age group. Therefore, BCS development may be attracting more young developers than the veterans. 

 In terms of the highest level of education (Figure~\ref{fig-RQ1}(c)), our respondents are more qualified with 15.4\% high school graduate, 46.8\% with a bachelors degree, 32.7\% with a masters degree, and 5.1\% with a Ph.D. The corresponding numbers in the SO survey are 23.4\%, 47.7\%,  23.2\%, and 2.2\% respectively. The higher educational qualifications of our respondents may not be surprising as BCS development requires more in-depth knowledge of computing than non-BCS development~\cite{zheng2016blockchain}.
 
In terms of software development experiences (Figure~\ref{fig-RQ1}(d)), 70.5\% of our respondents have more than five years of development experiences and 42.3\% have more than 10 years. The corresponding numbers in the SO survey are 62.3\% and 34.1\% respectively. It indicates that the BCS developers are likely to be more experienced in software development than their non-BCS counterparts. 

\begin{boxedtext}
\textbf{Key takeaway 1: } In general, BCS developer population is more qualitifed than the general software developer population. Although, we noticed that more than 81\% of our respondents have less than 2 years of BCS development experience, more than 70\% developers from the same group were found to have more than five years of development experiences.
These numbers indicate that a large number of software developers, who are experienced in non-BCS development, have recently joined BCS projects potentially due to the recent hypes generated by the blockchain technology.
\end{boxedtext}

\subsection{RQ2: Motivations of BCS Developers}
\label{sec-RQ2}

\begin{figure}
\includegraphics[width=0.8\linewidth]{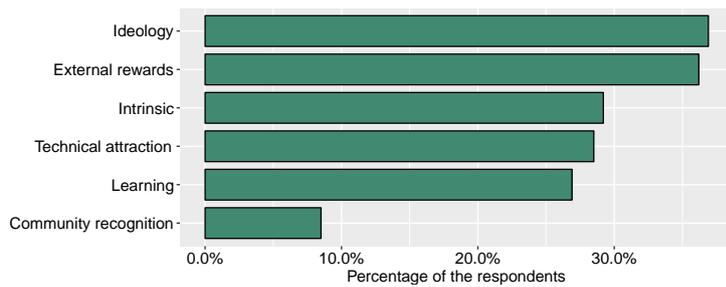}
\caption{Primary motivations of BCS developers'}
\label{fig-motivations}
\end{figure}

Figure~\ref{fig-motivations} shows the primary motivations of BCS developers, which emerged from the answers to Q10 (Table~\ref{table:survey_questions}) of our survey. 
Besides \textbf{technical attraction}, the other five categories  of motivations that BCS developers have, are similar to  the motivations of Open Source Software (OSS) developers~\cite{hars2001working,lakhani2003hackers,ye2003toward}. Since blockchain is a new technology, many BCS developers indicate their fascinations to this innovative technology as one of their primary motives. While \textbf{ideology} did not top the list of OSS developers' motives~\cite{hars2001working,lakhani2003hackers,ye2003toward}, it ranks top for the  BCS developers.

\begin{figure}
\centering \includegraphics[width=\linewidth]{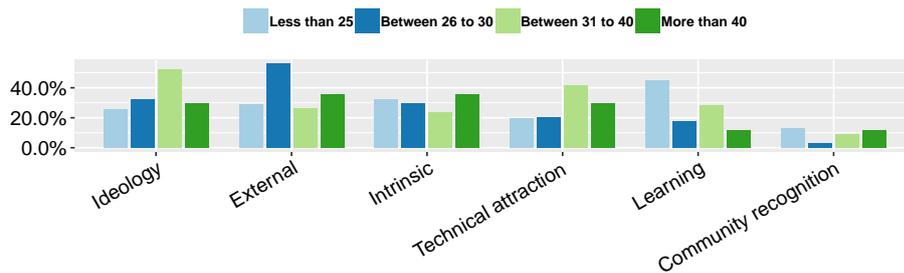}
\caption{Age vs. Motivation (percentage computed based on motivations within the same age group)}
\label{fig:age-vs-motive}
\end{figure}

Since results from the  psychology domain suggest that a person's motivation may vary based on his/her age, education level~\cite{hitka2015impact}, compensation~\cite{hennessey1998reality}, or gender~\cite{meece2006gender}, we investigated whether those factors had any impact on a BCS developer's motives. 
Our results suggest significant differences (Chi Square,  after applying False Discover Rate (FDR) corrections~\cite{benjamini1995controlling} for multiple comparisons) in motivations based on the age ($\chi ^2 =33.34, p=0.03$) or compensation ($ \chi ^2 = 40.58, p=0.01$) of a respondent but no significant difference was found based on the level of education ($ \chi ^2 = 22.40, p=0.29$) or gender ($ \chi ^2 = 9.83, p=1.0$). While most of the BCS developers between 31 to 40 years of age have ideological motives, developers aged between 26 to 30 years are more likely to be motivated by external rewards and developers aged 25 years and younger are more likely to be motivated by the prospects of learning (Figure~\ref{fig:age-vs-motive}). On the other hand, developers receiving some types of direct compensations are more likely motivated by external rewards.
The following subsections examine these motivations in more detail.

\subsubsection{Ideology }
\label{subsec-ideology}
The primary motivation behind Bitcoin, the first blockchain based cryptocurrency was to create a decentralized currency that cannot be manipulated by a central authority. 
More than one-third of our respondents (36.9\%) are motivated by a similar ideology.

\surveyquote{I truly believe in the right to have a private way to send and store money. Also removing power from banks and governments. [\#195]}

\subsubsection{External rewards }
Some developers (36.2\%) contribute to BCS projects to earn money either by working part-time or by accepting bounty offers. Developers, who hold cryptocurrency, are naturally motivated to increase its value.
\surveyquote{.. earn money; I get salary, and also I bought coins so their growth will give me money. [\#147]}

Many developers are full-time employees of the organization that manages his/her project.
\surveyquote{I am paid to work full-time contributing to blockchain projects. [\#75]}

\subsubsection{Intrinsic}
Due to the various programming challenges that BCS development offers, many developers (29.2\%), who enjoy writing code to solve problems, contribute to BCS projects.

\surveyquote{I love coding. I love to solve problems and support people. [\#127]}
Some contribute to BCS projects due to their passions for a particular area.

\surveyquote{It's an opportunity to work on a programming language, which I've always wanted to do. [\#142]}

\subsubsection{Technical attraction }
Attraction to the blockchain technology is one of the motivations for many BCS developers (28.5\%).
\surveyquote{Curiosity mainly. I was studying cryptography by myself before that and saw a chance to see it applied in unbelievable ways. [\#31]}

\subsubsection{Learning }
 Developers (26.9\%) considering blockchain as a promising technology for the future want to learn and add it to their portfolio. 
\surveyquote{... to develop a better understanding of making and working of a blockchain. [\#103]}

\subsubsection{Community recognition}
Some respondents (8.5\%) want to improve their portfolio through their involvement and recognition in the BCS community.
\surveyquote{.. become more famous in the community (get good reputation). [\#63]}

\begin{boxedtext}
\textbf{Key takeaway 2:} Due to the significant financial gains by the early cryptocurrency investors as well as a large influx of cash through ICOs, we hypothesized that the majority of the BCS developers might be motivated by external rewards. However, the ratio of our respondents reporting external motives (36\%) were similar to the number from prior OSS studies~\cite{hars2001working,lakhani2003hackers}. Moreover, the ratio of volunteers (43\%) is also similar to what was reported in prior studies~\cite{lakhani2003hackers,Bosu-etal:JSS}. On the other hand, ideological motives were more frequent among  BCS developers (37\%) than among the OSS developers~\cite{hars2001working,lakhani2003hackers}. Therefore, getting aligned with the \textit{ideology} of a BCS community is important to become a member of that community.
\end{boxedtext}

\subsection{RQ3: Differences Between BCS and non-BCS Development}
\label{sec-RQ3}
\begin{figure}
\centering \includegraphics[width=0.9\linewidth]{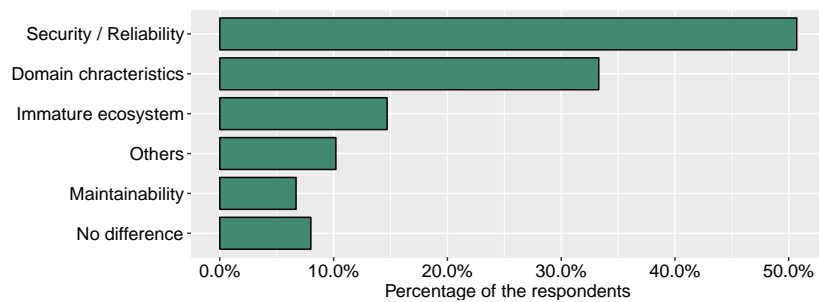}
\caption{Differences between BCS and non-BCS development}
\label{fig-practices}
\end{figure}

Our survey included two questions to find the differences between BCS and non-BCS development.  
Figure~\ref{fig-practices} shows the primary differences between  BCS and non-BCS development as indicated by the respondents of our survey in response to Q11. 
Although BCS development has many similarities to traditional OSS development, due to some unique characteristics of the BCS domain, 93\% of our respondents reported BCS development as somewhat different from non-BCS development. 
Majority of our respondents did not encounter such strict security or reliability requirements while developing a non-BCS. The unique characteristics of the BCS domain (e.g., immutability, difficulty in upgrading the software) was also a differentiating factor for more than one-third respondents.
 
The following six subsections detail those differences. We conclude this section with a comparison of BCS development with three software development domains from Microsoft (Section~\ref{industry-comparison}).

\subsubsection{Security/reliability}
\label{subsec-security}
A higher emphasis on security and reliability is the primary factor that differentiates BCS from most of the non-BCS.
Since the primary applications of the blockchain technology are maintaining ledgers of financial transactions,  ensuring the security and reliability of a BCS application is the highest priority for majority of the BCS developers (50.7\%).
\surveyquote{Security and backward compatibility are held with utmost importance here unlike some other FOSS projects. [\#3]}

A single defect in a BCS application can cost millions of dollars. 
For example, recently a  bug in the parity wallet code enabled hackers to steal \$30M worth Ethereum tokens~\cite{santiagopalladino2017}.
\surveyquote{..you should be more careful because there is too much at stake (nowadays a lot of money are invested in cryptocurrencies). In most of the projects (non-blockchain) when a bug appears, it will be fixed and soon forgotten. But in blockchain projects some bugs can be very costly and never forgotten. [\#147]}

\subsubsection{ Domain characteristics}
\label{subsec-domain-diff}
One-third of our respondents (33.3\%) consider several characteristics of the BCS domain as factors differentiating BCS development from non-BCS.
First, data stored in a blockchain is immutable. In other domains, there are several mechanisms to fix errors later by altering data. However, altering a blockchain ledger is almost impossible. 
\surveyquote{... also a disadvantage that you can not fix problems due to human errors or bugs in that transactions can not be changed. [\#97]}

Second, compared to the most non-BCS applications  that operate on centralized and/or hosted environments, BCS applications operate on a complex, secured, distributed and decentralized network.
\surveyquote{The distributed nature of blockchain software development makes it difficult to build robust software. Unreliable connections, unexpected latency, and malicious nodes create a hostile production environment. [\#135]}

Third, blockchains use public key cryptography and cryptographic hash functions to store and verify transactions. Cryptography is difficult to master and very few other domains require similar in-depth knowledge of cryptography as the BCS domain. Moreover, the knowledge of networking and networking security is a must for BCS development.  The daunting requirement of having knowledge about diverse technological areas is another differentiating factor.
\surveyquote{All kinds of knowledge are involved. It's hard to comprehend the whole project, including cryptography, network programming, economy policy etc.[\#166]}

Finally, the blockchain technology is changing rapidly with new protocols, innovations, and possibilities emerging everyday. BCS projects that are not able to evolve rapidly are at the risk of loosing their market capitalization. 
\surveyquote{Technology moves fast and blockchain software development (in general, not just Ethereum) is moving at an extremely fast pace. Need to keep up and adjust to meet whatever new requirements arise.[\#98]}

\subsubsection{Immature ecosystem}
Many of the innovative aspects of the blockchain technology (e.g., smart-contract, privacy) are relatively new. Although the number of BCS projects have grown exponentially during the last couple of years, many tools and libraries  that may support BCS development are still missing. Even the tools that currently exist are not stable. Therefore, many respondents (14.7\%) consider the BCS  development ecosystem as immatured compared to most of the non-BCS counterparts.
\surveyquote{We are in 1986 on a web development timeline, to my mind.  Like everything is in C can hit memory leaks, crashes, etc.  Everything very raw and dangerous.   We still have most of the stack to build. [\#137]}

Since Blockchain is a new technology, there is scarcity of enough domain-experienced developers compared to most non-BCS domains. The demographics of our survey also shows  more than 83\% respondents with less than two years of experience in BCS development. 
\surveyquote{People have more experience in other areas. [\#48]}

\begin{landscape}
\begin{table*}
	\centering
    \caption{Comparison of BCS development with three non-BCS domains at Microsoft. \colorbox{gray!35}{Grey}background indicates a statistically significant difference. For the effect sizes ($r$), a value in \textcolor{red}{red} indicates BCS developers agreeing less with that statement,  while a value in \textcolor{dgreen}{green} indicates more agreement from them. If an effect size is statistically significant, the background color intensity of the cell indicates strength of the effect size with darker shades indicating larger effect sizes. For example, negative effects are presented by: i) \colorbox{orange!25}{small}, \colorbox{orange!50}{medium}, and \colorbox{orange!75}{large} and positive effects are presented by: i) \colorbox{blue!10}{small}, \colorbox{blue!20}{medium}, and \colorbox{blue!30}{large}.  The background color Statistical significance (p) is reported after applying correction for multiple comparisons using False Discovery Rate (FDR)~\cite{benjamini1995controlling}. }
	\label{table:comparison}
\resizebox{\linewidth}{!} {
\begin{tabular}{|l|p{7.5cm}|p{1.7cm}|r|r|r|r|r|r|}
\hhline{---------}
&\multirow{2}{*}{\textbf{Statement}}  &  \multirow{2}{1.7cm}{\textbf{Rating\newline distribution}} & \multicolumn{2}{c}{\textbf{BCS Vs. Games}} & \multicolumn{2}{|c}{\textbf{BCS Vs. Office}} & \multicolumn{2}{|c|}{\textbf{BCS Vs. Other}} \\ \hhline{~~~------}
& & & $p$  & $r$ & $p$ & $r$  &$p$ & $r$ 
\\ \hhline{-|-|-|-|-|-|-|-|-|}

\multicolumn{9}{l}{ \hspace*{1cm}\textbf{ A. Quality assurance}}\\ \hline
QA1.& My software is well tested manually (e.g., paid testers thoroughly use the software).&  \ticker{.9}{11}{30.3}{41.3}{16.5}
 & 0.82 &\moren{ 0.06} &   0.052 & \lessn{ 0.14} &   0.09 &  \moren{0.12} \\ \hhline{-|-|-|-|-|-|-|-|-|}
QA2.& My software is well tested by automated testing (e.g., scripts that thoroughly use the software).&  \ticker{2.8}{8.3}{19.3}{45.9}{23.9}
 &  \significant{ $<0.01$ }& \more{ 0.32} & 0.09 & \moren{ 0.12} &  \significant{ 0.02} &\moreSmall{ 0.18} 
 \\ \hhline{-|-|-|-|-|-|-|-|-|}
QA3.& My software is well tested by unit tests.&  \ticker{2.8}{11.0}{18.3}{47.7}{20.2}
 &  \significant{ $<0.01$} &\more{ 0.34} & \significant{ $<0.01$} & \more{0.31} & 0.57 &\moren{ 0.01} \\ \hhline{-|-|-|-|-|-|-|-|-|}
QA4.& Bugs in my software are hard to diagnose. &  \ticker{4.6}{26.6}{33.0}{32.1}{3.7}
&  \significant{ $<0.01$} & \moreSmall{ 0.23} & 0.19 & \moren{ 0.08} & .34 & \moren{ 0.06} \\ \hhline{-|-|-|-|-|-|-|-|-|}

QA5.& It's difficult to write thorough automated tests for my software because it's so complex.&   \ticker{8.3}{34.3}{27.8}{25}{4.6}
 &  \significant{ 0.02} & \lessSmall{ 0.15} & 0.13 &\lessn{ 0.1} & 0.93 & \lessn{0.1} \\ \hhline{-|-|-|-|-|-|-|-|-|}


\multicolumn{9}{l}{\hspace*{1cm}\textbf{B. Development process}}\\ \hline
DP1.& My team has flexible release deadlines.&  \ticker{3.7}{8.3}{17.4}{45.9}{24.8}
 &  \significant{ $<0.01$} & \more{ 0.46} &  \significant{ $<0.01$} &\moreLarge{ 0.53} &   \significant{ $<0.01$} &\more{ 0.44}
\\ \hhline{-|-|-|-|-|-|-|-|-|}
DP2.& My software has clear functional requirements.&  \ticker{1.8}{7.3}{21.8}{49.1}{20}
 & \significant{ 0.02} & \moreSmall{ 0.14} & 0.64 &\lessn{ 0.02} & 0.93 &\moren{ 0.09} \\ \hhline{-|-|-|-|-|-|-|-|-|}
 DP3. &My team uses a waterfall process, rather than an agile process.&  \ticker{30}{32.7}{30}{7.3}{0}
 &   \significant{ 0.04} & \moreSmall{ 0.12} & 0.65 &\lessn{ 0.03} &  0.09 & \lessn{ 0.08} \\ \hhline{---------}
DP4.& My team adheres strictly to a process (for example, scrum or waterfall).&  \ticker{19.3}{33.9}{26.6}{15.6}{4.6}
 &  \significant{ $<0.01$} & \lessSmall{ 0.2} &  \significant{ $<0.01$} & \less{ 0.37} & \significant{ $<0.01$} & \lessSmall{ 0.29} \\ \hhline{---------}
 
 
 \multicolumn{9}{l}{\hspace*{1cm}\textbf{C. Software maintenance}}\\ \hline
 SM1.& My software has high technical debt (for example, a lot of hacks). &  \ticker{19.1}{40}{26.4}{10.9}{3.6} 
 & 0.3 &\lessn{ 0.04} & 0.13 &\lessn{ 0.1} & 0.53 &\lessn{ 0.02} \\ \hhline{-|-|-|-|-|-|-|-|-|}
SM2.& The technical debt is likely to be paid down in the future (for example, through refactoring).&  \ticker{1.8}{13.6}{29.1}{41.8}{13.6}
 &  \significant{ $<0.01$} &\moreSmall{ 0.18} & \significant{ $<0.01$} & \moreSmall{ 0.23} &  \significant{ 0.02} & \moreSmall{ 0.17} \\ \hhline{-|-|-|-|-|-|-|-|-|}
 SM3.& My software's architecture evolves significantly as the software gets more mature.&  \ticker{.9}{9.1}{15.5}{55.5}{19.1}
 & 0.26 & \moren{ 0.05} &   0.06 & \moren{ 0.09} & 0.51 & \moren{ 0.03} \\ \hhline{-|-|-|-|-|-|-|-|-|}
 \multicolumn{9}{l}{\hspace*{1cm}\textbf{D. Ideology / morale} }\\ \hline
 DM1.& My software creates value for society.&  \ticker{0}{0}{15.7}{40.7}{43.5}
 &  \significant{ $<0.01$} &\moreSmall{ 0.26} & 0.21 &\moren{ 0.07} &  \significant{ 0.02} &\moreSmall{ 0.17} \\ \hhline{-|-|-|-|-|-|-|-|-|}
DM2.& Creativity is highly valued on my team.&  \ticker{.9}{3.7}{25.7}{43.1}{26.6}
 &  \significant{ 0.03} & \lessSmall{ 0.14} & 0.48 & \moren{0.02} & 0.53 &\moren{ 0.02} \\ \hhline{-|-|-|-|-|-|-|-|-|}
 
DM3.& Creating my software is challenging.&  \ticker{.9}{8.3}{13.9}{54.6}{22.2}
 & 0.24 & \lessn{ 0.06} & 0.56 &\lessn{ 0} & 0.74 & \lessn{0.03} \\ \hhline{-|-|-|-|-|-|-|-|-|}
DM4.& When I tell people outside of my company about the software I work on, they are impressed.&  \ticker{.9}{1.9}{24.1}{50.9}{22.2}
 & \significant{ 0.04} &\lessSmall{ 0.13} & 0.13& \moren{0.1} &  0.09 &\moren{ 0.11} \\ \hhline{-|-|-|-|-|-|-|-|-|}

\end{tabular} 
}
\end{table*}
\end{landscape}
\subsubsection{Maintainability}
\label{subsec-maintain}
According to 6.7\% of our respondents, maintaining a BCS application is difficult  compared to a non-BCS application.
Blockchains usually incorporate new functionality through hard forks~\cite{hardfork}, which is usually scheduled way ahead of time to ensure that all the nodes are running the same version of the software. Due to the difficulty in upgrade, a BCS application needs to be well tested before a release.
\surveyquote{.. non-blockchain software can easily upgrade features but blockchain software needs to wait for the preparation of nodes all over the world to upgrade functionality. [\#22] }

BCS applications run on public blockchains that have thousands of blocks created over the years. Therefore, a BCS application must be backward compatible  and capable of validating earlier transactions or executing smart-contracts deployed through earlier versions.
\surveyquote{Ethereum is a public chain. So we have to consider backward compatibility all the time. [\#28]}

\subsubsection{No difference}
\label{sec-no-diff}
Some BCS developers (8\%) do not find much difference between  BCS and non-BCS development.
\surveyquote{Not that different from other performance-sensitive, high reliability software. [\#9]}

\subsubsection{How does BCS development differ from software development at Microsoft?}
\label{industry-comparison}
The `Rating distribution' column in Table~\ref{table:comparison} shows the distributions of the  ratings by our respondents for 16 statements ( The `Statement' column in Table~\ref{table:comparison}). The leftmost bar indicates `Strongly disagree,' the middle bar indicates `Neutral,' and the rightmost bar indicates `Strongly agree.' We do not plot the ratings from the Microsoft developers since a prior publication reports those~\cite{murphy2014cowboys}.

Since the Shapiro-Wilk~\cite{shapiro1965} test indicates that those ratings significantly differ from a normal distribution, we use non-parametric statistics.  For each of the statements, we compare the responses from BCS developers with the responses of Microsoft developers from three domains (i.e., Games, Office, and Others). The three `$p$' columns show statistical significance of the differences based on the Mann-Whitney U test after applying corrections for multiple comparisons using the FDR method~\cite{benjamini1995controlling}. The three `$r$' columns show the effect sizes estimated using Rosenthal's formula~\cite{rosenthal1994parametric}. Rosenthal~\cite{rosenthal1996qualitative} also recommends interpreting those effect sizes as: $r\geq 0.5\Rightarrow$ \emph{large}, $r\geq 0.3\Rightarrow$ \emph{medium}, and $r\geq 0.1\Rightarrow$ \emph{small}.

Posthoc, we grouped the 16 statements  into following four categories based on the theme of each statement.

\vspace{3pt}\noindent \textit{A. Quality Assurance:}
BCS developers agree more than both  \emph{Games} and \emph{Office} developers that their software is well tested using unit tests (QA3). They also consider BCS more thoroughly automated tested (QA2) than both \emph{Games} and \emph{Other}. However, for manual testing (QA1), we did not find any significant difference. From the remaining two statements, BCS developers agree more than the  \emph{Games} developers on bug diagnosis difficulty (QA4)  but disagree with them on the difficulty of writing unit tests (QA5). These results indicate BCS developers' higher emphasis on automated testing to ensure the quality of their software.

\vspace{3pt} \noindent \textit{B. Development process:}
Among the four groups, BCS developers have the most flexible deadlines (DP1). Since the costs of defects are very high and patching a defect after release is very difficult, BCS developers are more flexible in postponing a release.
 On the other hand, among the four groups they are the least strict in following a  process (DP4).

\vspace{3pt}\noindent \textit{C. Software maintenance:}
Among the four groups, BCS developers are the most focused on paying technical debts (SM2), which further emphasizes the importance of long term maintainability of a BCS application. For the remaining two statements the differences were not significant. 

\vspace{3pt} \noindent  \textit{D. Ideology/morale:}
In terms of ideology/morale, BCS has no significant difference with \emph{Office}. The results of comparisons with the remaining two groups are mixed. On taking pride outside, for their work (DM4), BCS developers agree less than \emph{Games} but are not  significantly different than \emph{Other};  while on creating value for the society (DM1), BCS developers top both. BCS developers also agree less than  the \emph{Games} developer on the valuation of creativity (DM2) within their team.

In summary, BCS has the most differences with \emph{Games} (i.e., 12), followed by \emph{Other} (i.e., 5) and the least with \emph{Office} (i.e., 4). BCS development has higher emphasis on security, reliability, and maintainability but lower emphasis on following a process or deadline than the other three domains. Apart from those differences, BCS may not be much different from traditional development domains, especially widely used software, such as \emph{Office}.

\begin{boxedtext}
\textbf{Key takeaway 3:}
So, is BCS development really different?
The answer to this question will depend on whom we ask. It's true that BCS development has a very high emphasis on security and reliability, but many of the existing software development domains (e.g., financial transaction, air traffic controller, and nuclear power plant management) have a similar emphasis on security and reliability.   If a developer's non-BCS experience is in high assurance software (Section~\ref{sec-no-diff}), then he/she might find little differences. However, over 93\% of our respondents' non-BCS experiences significantly differed from their experiences in BCS (Figure~\ref{fig-practices}).  Our survey received responses from $\approx$10\% of the most active developers from the top BCS projects, and 70\% of our respondents have more than 5 years of development experiences (Section~\ref{sec-RQ1}). Yet, they found BCS development different from non-BCS. Some of the differences, such as the immaturity of the ecosystem, will resolve with time, but others, such as immutability of data as well as difficulty in upgrading the software after deployment, which is rare among the non-BCS domains, will linger as a differentiating factor. 
\end{boxedtext}

\subsection{RQ4: Challenges of BCS Development}
\label{sec-RQ4}
\begin{figure}
\centering
\includegraphics[width=0.9\linewidth]{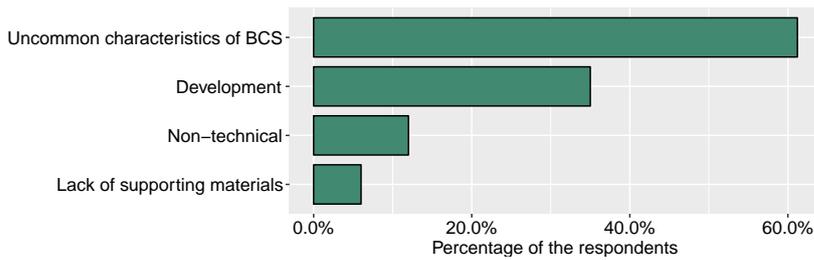}

 \caption{Primary challenges of BCS development}
\label{fig-challenges}
\end{figure}

Figure~\ref{fig-challenges} shows the primary types of challenges of BCS development according to the respondents of our survey (Q13: Table~\ref{table:survey_questions}).
Since prior results from the Psychology domain suggest that a person's ability to acquire knowledge of a new domain depends on his/her age and education level~\cite{beier2005age}, we investigated whether those factors have any impact on the challenges reported by the BCS developers from our survey. Although we did not find any differences based on the education level ($ \chi ^2 = 14.46, p=0.27$), we found that the developers aged 40 years and older, who are likely to have significant experiences in non-BCS domains,  were more likely ($\chi^2= 35.07, p=0.02$) to be challenged by the characteristics of the BCS domain (Section~\ref{sec-bcs-domain}) than the other age groups. 

\subsubsection{Uncommon characteristics of BCS}
\label{sec-bcs-domain}

Some of the characteristics of the BCS domain that are rare among non-BCS, are sources of challenges for more than 61\% BCS developers. Around 70\% respondents of our survey have more than five years of software development experience.  
Therefore, the aspects of BCS development that differ from a non-BCS are primary  sources of challenges for them.
While discussing BCS development challenges, our respondents reiterated following differences already mentioned earlier (section \ref{sec-RQ3}):
\begin{itemize}
\item high costs of defects (section~\ref{subsec-security});
\item technological complexity (section~\ref{subsec-domain-diff});
\item distributed, decentralized, and hostile environment (section~\ref{subsec-domain-diff});
\item rapidly changing ecosystem (section~\ref{subsec-domain-diff}); and
\item difficulty in maintenance (section~\ref{subsec-maintain}).

\end{itemize}

Steep learning curves to get familiar with a BCS project is another challenge for many BCS developers. The codebase of a BCS project is not only complex but also requires a sound understanding of cryptography, networking, distributed systems as well as project specific protocols. 

\surveyquote{The bitcoin core codebase is an incredibly complicated system, so the hardest part is building up a good enough understanding of it such that it's safe to make changes. [\#156]}

\subsubsection{Development}

Development challenges are related to testing, ensuring security, and reviewing code as mentioned by 35\% of our respondents.
Testing blockchain software is challenging as the software executes on a distributed and potentially hostile environment that  currently cannot be adequately simulated on a development machine.
\surveyquote{.. to test that bugs are not included in important parts such as consensus. [\#22]}

The security of a BCS software is the highest priority as it handles financial data that can be exploited for financial gains. Therefore, BCS developers must consider security aspects when writing code.

\surveyquote{... thinking with a security mindset. the software is literally money, so when writing it you have to be wary of any way in which things could go wrong.[\#143]}

Due to the lack of quality reviewers, open source software (OSS) developers often have to wait for a long time to get their code reviewed~\cite{bosu2014impact}. BCS developers also  report similar challenges. 
\surveyquote{... it is often difficult to get code reviews on open source projects even with a history of contributions. [\#75]}

\subsubsection{Non-technical}

The non-technical challenges of BCS development, which were reported by 12\% of our respndents, are due to collaboration issues, difficulties in reaching agreement among the community members, and the ethical aspects of BCS. 
Most of the BCS projects are run by communities. However, many of the projects have raised over hundreds of millions of dollars through initial coin offerrings~\cite{howell2018initial} and have very high valuations~\cite{coinmarketcap}. Since a large sum of money is at stake, community members often engage in arguments to decide a project roadmap~\cite{bitcoin-poitics}.
\surveyquote{Governance, deciding on changes, and set a future roadmap. [\#35]}

The development team comprising volunteers with diverse personalities, background, experience, and motivation often encounter issues to collaborate.
\surveyquote{nothing particular to the blockchain, prickly personalities! [\#51]}

Some BCS developers are primarily motivated by the prospects of financial gains and will not hesitate to adopt an unethical approach if presented with an opportunity. Without a central monitoring authority, preventing developers with malicious intents is a challenge.
\surveyquote{human greed, and cleaning up the mess after bounties and scams on forum.[\#44]}

\subsubsection{Lack of supporting materials}
\label{sec-no-learning}
 The lack of supporting tools and documentation is a source of challenges for many BCS developers (6\%).
 Many of the supporting tools that can help their development tasks are yet to be developed. Moreover, the tools that are currently available are immature and unreliable. 
 \surveyquote{Reliability of and the lack of good development tools.  Like testing frameworks, and the difficulty of debugging.[\#16]}
To understand a complicated domain such as BCS, developers are looking for good learning materials and  tutorials, but such materials are currently rarities. Even the documentation that they currently have are not user friendly.

\surveyquote{Blockchain is totally a new technology so there is few information we can find.}

\begin{boxedtext}
\textbf{Key takeaway 4:}
Since most of the non-BCS domains do not have similar high reliability and security requirements as the BCS domain, developers coming from other domains (except high assurance software) will encounter challenges due to those differences. Moreover,
BCS Developers must be careful in writing code due to high costs of defects as well as difficulty in upgrading the software. Yet, they are under constant pressure due to a rapidly changing ecosystem as well as high expectations from the stakeholders to release new versions. Moreover, as the blockchains become more popular the scalability of BCS software has become an area of concern~\cite{porru2017blockchain}. As a result BCS development is challenging even for developers with considerable non-BCS experiences (Section~\ref{sec-bcs-domain}).

\end{boxedtext}

\subsection{RQ5: Tools that BCS Developers Need}
\label{sec-RQ5}
\begin{figure}
\centering \includegraphics[width=0.9\linewidth]{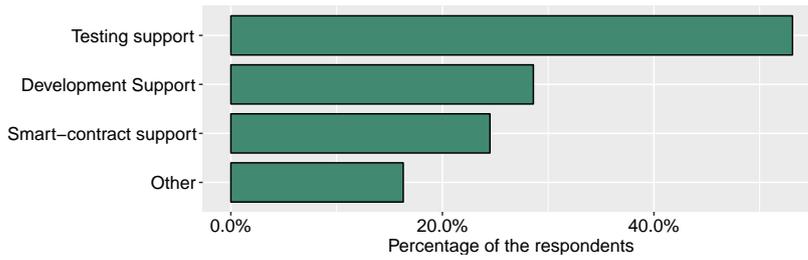}
\caption{Categories of tools that BCS developers need}
\label{fig-tools}
\end{figure}

We asked the respondent of our survey to describe  the type of tools that they currently do not have but if implemented can significantly improve their development productivity (Q14: Table~\ref{table:survey_questions}). In response, our respondents indicated their needs for four categories of tools (Figure~\ref{fig-tools}).  We also hypothesized the the requirement for supporting tool may vary based on the software development experience or BCS development experience of a respondent. However, the results of our analysis did not find any statistically significant difference.

\subsubsection{Testing support}
\label{sec-testing-support}
Majority of our respondents (53.1\%) suggest that \emph{testing} is the area of BCS development  that currently needs supporting tools the most. BCS developers use `testnets'~\cite{testnet} to experiment their code before deploying it to the `mainnet'~\cite{mainnet}. Many developers experience difficulties in setting up a testnet. An easy to setup, `one-click testnet' may be a solution. 
\surveyquote{Easy way of forking mainnets for testing purposes, a way to deploy a test net in one click would be nice.[\#150]}

The simulators that BCS developers currently have are limited and are unable to simulate a   complex and hostile real world environment. While they have testnets, but the scales and complexities of testnets are no where closer to a mainnet. Few recent works~\cite{stoykov2017vibes,chen2017using} have attempted to build such simulators, but no satisfactory solution exists yet.
\surveyquote{Easier ways to simulate complex network topologies on one single machine to simulate the network. [\#195]}

Formal verification~\cite{clarke1996formal} techniques have been useful to secure BCS projects. However, developers find current formal specification languages (e.g., TLA+, VDM, and Z-notation) very complex to learn and use. They wish for user friendly alternatives. 
\surveyquote{End-to-end formal specification and verification tools with notations that mere programmers can understand.  I sometimes think our formalists actually like to make their work obscure. [\#1]}

Static analysis and penetration testing tools have been useful in non-BCS domains for security testing~\cite{chess2004static,arkin2005software}. Since, those tools do not work well on BCS codebase, developers wish for automated security testing tools designed specifically for the BCS domain.
\surveyquote{Fuzz testing, something like linting for security best practices. [\#196]}

\subsubsection{Development support}
The IDEs, designed for the non-BCS domains, lack adequate supports for testing and debugging a BCS codebase. Therefore, BCS developers use an array of tools for various development activities.
Some developers (28.6\%) maintain that an IDE designed specifically for BCS development would help them.
\surveyquote{... they are all mostly disconnected, I have to switch from one tool/platform to another for my regular development activities many times every day. It would help if there were better integration between them all. [\#66]}

Even the tools that exist are not reliable, and BCS developers wish them to be more stable.
\surveyquote{Tools exist that are in their infancy and just need to get better. [\#13]}

\subsubsection{Smart-contract development}
\label{sec-smart-contract}
Smart-contracts are written using contract-oriented programming language such as Solidity or Vyper and then compiled into bytecode for a platform (e.g., Ethereum Virual Machine aka EVM). Remix, an IDE for solidity, currently lacks many features such as: error highlighting and line by line debugging, that many BCS developers (24.5\%) wish for. 
\surveyquote{Better tool support for smart-contract development. IDE integration with interactive debugging. [\#137]}

Before interacting with a smart-contract, a developer might want to verify its security properties by decompiling its bytecode. While few solutions exist~\cite{suiche2017porosity}, smart-contract developers wish for a reliable and user-friendly decompiler.

\surveyquote{... high-level Solidity decompiler that works (the current EVM-to-Solidity decompilers are horrible). [\#113]}
\subsubsection{Others}
The other tools wished by the respondents include UML/design notations for the BCS domain, containers for deployment, and automated performance analysis tools.

\begin{boxedtext}
\textbf{Key takeaway 5:}
Based on the personal experiences of our respondents, they found some widely used tools tuned for non-BCS development, lacking required support for BCS development. While some of the needs expressed by our respondents (e.g., easy to write formal specification) may be a wishful thinking and difficult to achieve, most of those tools are feasible.  Potentially implementable tools for BCS development include: testing environment,  automatic security testing, static analysis for smart-contrats, and easy to deploy testnets. 
Since an array of research predict tremendous impacts of the blockchain technology and smartcontracts in future~\cite{iansiti2017truth,peters2016understanding,fanning2016blockchain}, the number of BCS projects and developers contributing to those projects will grow significantly over the next decade. Therefore, research and development efforts should focus on implementing those tools to build a mature BCS development ecosystem. 
\end{boxedtext}
\section{Implications}
\label{sec-discussion}
In the following subsections, we discuss key  implications of our results.

\subsection{Notes for Prospective Joiners}
\emph{Ideology} drives more than one-third of the BCS developers (Section~\ref{subsec-ideology}). Prior research suggest that OSS developers who do not share a common ideology with the community are not only less productive~\cite{stewart2006impact} but also fail to form synergies with the existing members. There are several instances where ideological conflicts have split a BCS community into two rival projects (e.g., Ethereum $\rightarrow$ Ethereum and Ethereum Classic). Therefore, a prospective joiner should select a project that has ideologies aligned with his/her beliefs.

Second, an experienced non-BCS developer, regardless of his/her prior non-BCS domain, will encounter differences due to several unique characteristics of the BCS domain (Section~\ref{subsec-domain-diff}). A prospective joiner must be prepared to conduct rigorous testing to ensure the correctness of code before deployment, since-- i) data cannot be altered once written on the blockchain and ii) updating BCS is very difficult compared to non-BCS. Despite such extensive correctness requirements, tools and frameworks to conduct such rigorous tests are either limited or non-existent (Section~\ref{sec-testing-support}). Therefore, a prospective joiner must be ready to spend significant manual efforts to test BCS for correctness, as our respondent\#29  stated, \textit{``Testing is 80\% of development.''}

Finally, BCS has steep learning curve with knowledge requirements in Cryptography, networking, security and distributed computing (Section~\ref{sec-bcs-domain}). Yet, documentations and tutorials to assist the learning process of a newcomer are limited (Section~\ref{sec-no-learning}). Therefore, a BCS joiner needs to be prepared to spend considerable efforts researching blockchain concepts and to utilize community Q\&A sites (e.g., \url{https://bitcoin.stackexchange.com}).

\subsection{Education and Preparation}
Our results suggest that special skills, beyond those required for most non-BCS development would be beneficial for persons considering to join BCS. Chief among them is the knowledge of secured programming (Section~\ref{subsec-security}). Unlike the majority of non-BCS, every module of a BCS is subject to attacks by malicious actors, since the stakes are substantially higher (Section~\ref{subsec-domain-diff}). Second, due to its distributed and untrusted environment,  a BCS developer needs in-depth networking knowledge to design and secure communication between distributed nodes (Section~\ref{subsec-domain-diff}).
Third, a sound understanding of cryptography is required, since BCS is secured using encryption (Section~\ref{subsec-domain-diff}).
Finally, a lot of concepts of Blockchain are verified using principles of mathematics, therefore a sound math skill is essential to design new BCS algorithms or protocols (Section~\ref{subsec-domain-diff}).

\subsection{Suggestions for Tool Developers}
The study outcome also presents a requirement for developing an array of tools specialized for BCS development. As discussed in section~\ref{sec-RQ5}, potentially implementable tools for BCS development include: testing environment, debugging tools, automatic security testing, static analysis for smart-contracts, and easy to deploy testnets.

Since several characteristics of the BCS domain are different from a non-BCS domain, quite a few categories of tools (e.g., IDE, simulator, and static analysis tools), that are stable and mature for non-BCS development, either do not exist or lacks important support for BCS (Section~\ref{sec-testing-support}). Due to the lack of appropriate testing supports, BCS developers primarily depend on manual code reviews to ensure the security of their software~\cite{partha2018}. As a result, expensive bugs and hacks are very frequent in the BCS ecosystem~\cite{Parity,wan2017bug,porru2017blockchain}.

Smart-contracts, which open a new type of programming paradigm, are gaining popularities~\cite{luu2016making}. Yet, the smart-contract programming domain lacks basic development supports, such as an IDE with error highlighting, line by line debugging, and decompilers (Section~\ref{sec-smart-contract}). As a result, many of the smart-contracts deployed on public blockchains are vulnerable~\cite{nikolic2018finding}.

Finally, due to a higher emphasis on security and reliability, BCS developers are highly interested in writing formal specifications for their software. However, current formal verification languages and tools are difficult to learn and use for BCS (Section~\ref{sec-testing-support}). Therefore, another potential direction could be building customized formal specifications for the BCS domain.


\subsection{Research Directions}
The needs of BCS developers as identified in section~\ref{sec-RQ5} will provide guidance to  researchers to identify areas needing the most attentions.
One of such areas is testing, where BCS development would benefit from research efforts. While prior research has focused on testing distributed programs~\cite{campion2013system}, BCS operates on a different type of distributed environment where nodes are both collaborating and competing with each other other at the same. Moreover, some of those nodes may potentially have malicious intents. To simulate such networks  in a testing environment (Section~\ref{sec-testing-support}), research efforts are required to model their behavior. Such models may be also useful to mathematically verify the reliability and security of a network.

Second, smart-contracts automatically execute on distributed virtual-machines (e.g., EVM) based on certain predefined conditions. Since smart-contract programming languages (e.g., Solidity and Vyper) have different syntax and execution models, existing formal specification languages do not work well for those (Section~\ref{sec-smart-contract}). Although recent studies have focused on building new customized specifications to secure smart-contracts~\cite{hildenbrandt2018kevm,park2018formal}, more research effort is warranted as it is a high priority for the BCS developers.

Third, automated security testing is another key need for BCS developers. Fuzzing\footnote{providing invalid, unexpected, or random data as inputs}  technique has been successfully used for  automated security testing in various domains.  ContractFuzzer~\cite{jiang2018contractfuzzer}, a  recent tool, is the first fuzzer that aims to provide fuzzing support for Ethereum smart-contracts. However, such fuzzers are currently non-existent for other types of BCS. Research efforts to build fuzzers for various BCS would tremendously help the BCS ecosystem (Section~\ref{sec-testing-support}) as security is the top priority for BCS projects.
Finally, academic researchers, often in collaboration with practitioners, may focus on automated documentation generation, the lack of which is a barrier for the newcomers (Section~\ref{sec-no-learning}).




\section{Threats to validity}
\label{section-threats}
This section discusses the four common types threats to validity for this study.

\subsection{Internal validity}
\emph{Participant selection} is the primary threat to internal validity. We selected five commits as a threshold for an invitation to this survey. A  higher or lower threshold may have altered the results.  However, the results of our analysis did not find any significantly differences in opinions based on the number of pull requests of a respondent (Q8). Therefore, this threat may be minimal.

Although the response rate of our survey is similar to prior SE surveys~\cite{bosu2017process,kononenko2016code}, the response rate is only ($\approx$13\%). 
Therefore our results could suffer from a potential `non-response bias' (i.e., the opinions of the respondents who chose to participate may be different from who did not)~\cite{armstrong1977estimating}.  Even so, the 156 responses that we analyzed provide a rich source of data to reveal the insights described in this paper.

\subsection{Construct validity}

The primary threats to construct validity are related to the design of our survey. For example, respondents may misunderstand our questions or the questions may not be appropriate to investigate our research questions. Therefore, we took following measures during our survey design to reduce potential threats.

\begin{itemize}
    \item We conducted expert reviews as well as a pilot test to check both understandability and appropriateness of the survey questions.
    
    \item We carefully worded the open-ended questions in an unbiased manner.
    \item We provided clear instructions as well as asked the respondents to answer based on their personal experiences with BCS projects.
    \end{itemize}
 The actual responses our open-ended questions do indicate that the respondents understood the intent of the survey questions. Therefore, we think this threat is minimal.

Another potential threat to construct validity is our comparison of the BCS developers' ratings collected for this study with the ratings from the MS study.  On the comparability of two  different surveys, Herndon suggests that two surveys are comparable if  both surveys have:   i) the same target population, ii) the same objective, and iii) uses the same survey methodology~\cite{survey-link}. 
The MS study, empirically identified 28 statements spanning various software engineering as well as general work feature, to identify differences between game and non-game development. These statements were rated by three separate groups of developers to identify differences in  opinions among those groups. In this study, we have collected ratings from a fourth group (i.e. BCS developers) for 19 out of the 28 statements that are applicable to  BSC projects and compared that group with the three groups from the MS study. 
We believe data collected for our study is comparable with the MS study since both studies have:  i) software developers as the target population, ii) have the same objective of comparing the opinions of software developers from different domains, and iii) used a web-based survey to solicit ratings for a similar set of statements using the same Likert-scale. Moreover, comparing data from multiple surveys is not uncommon in many disciplines~\cite{elliott2005obtaining,sadana2002comparative,schenker2016combining,lohr2000inference,raghunathan2007combining}. Therefore, we consider this threat  to be minimal.

Finally, we have compared the demographics of our respondents against the demographics of the respondents reported in the SO survey. Comparisons of demographics across multiple surveys is common~\cite{mcdonald2007,Ross:2010,posel2006demographics}. Since our respondents come from the same population (i.e., software developers), a possible threat to validity for these comparisons is minimal.

\subsection{External validity}
The respondents of our survey may not adequately represent all BCS developers. While our respondents come from 61 different BCS projects, they primarily represent the top ones. Therefore, some of the opinions, especially, the challenges encountered by BCS developers in smaller projects may be different from those included in this study. 
However, our results also indicate that BCS developers' challenges are primarily due to the differences between BCS and non-BCS development, which may be similar across all projects.

Section~\ref{industry-comparison} compares BCS development with three non-BCS development domains from the same organization (i.e., Microsoft). Therefore, differences identified in that section may not apply to a different domain or a different organization.  A common misconception about industrial research at large companies such as Microsoft is that software projects at Microsoft are not representative of other software projects. While projects might be larger in size, most development practices at Microsoft are adapted from the general software engineering community and also used outside Microsoft. 
Since Microsoft is a well-recognized and mature software development organization, we believe that MS developers are eligible to provide us credible insights on non-BCS development,

\subsection{Conclusion validity}

We sent out the survey to 1,604 BCS developers that were eligible at that time. Using Yamane's~\cite{yamane1973statistics} formula for a recommended sample size, we would need $\approx$320 responses to obtain results that are within 5\% margin of error. Despite our best efforts, we obtained only 156 responses that are eligible for analyses. Therefore, many of our quantitative results are subject higher margin of errors. However, several prior SE surveys that have provided valuable insights to the community are also subject to similar errors as responses for SE surveys are generally low~\cite{Lee2017,kononenko2016code,bosu2017process}.

\section{Conclusion}
\label{sec-conclusion}
Despite the existence of a large number of active BCS projects as well as tremendous developer interests in the blockchain technology, there have been few empirical SE research exploring this area. In an attempt to bridge this gap, we studied the motivations, challenges, and needs of BCS developers. Our results suggest that 
although most of the BCS projects are Open Source Software (OSS) projects by nature, more than 93\% of our respondents found BCS development somewhat different from a non-BCS development as BCS projects have higher emphasis on security and reliability than most of the non-BCS projects. Other differences include:  higher costs of defects, decentralized and hostile environment, technological complexity,  and difficulty in upgrading the software after release. These differences were also the primary sources of challenges to the BCS developers. Software development tools that are tuned for non-BCS does not adequately support BCS development tasks and the ecosystem needs an array of new or improved tools, such as: a customized IDE for BCS development tasks, debuggers for smart-contracts, testing support, easily deployable simulators, and  BCS domain specific design notations. 

From the findings of our study, the prospective joiners of BCS development may gain some technical insight such as to learn secure programming practices, form mindset on test-centric development approach and also non-technical issues related to collaboration in the community of diverse motivation. Students wishing to join BCS should prepare themselves by gaining knowledge in networking, distributed programming,  cryptography, and mathematics. This study also identifies pressing need of tools supporting various BCS development tasks. Many of these tools require active participation of SE researchers who are expected to grab high-impact research problems from this research.

\begin{acknowledgements}
We are grateful to  Emerson Murphy-Hill from Google; Tom Zimmermann, and Nachi Nagappan from Microsoft Research for providing us the dataset of their Survey at Microsoft. We thank the anonymous reviewers for their thorough reviews and highly appreciate the comments and suggestions, which significantly contributed to improving the quality of this publication.
We thank the anonymous respondents of our survey.
\end{acknowledgements}

\bibliographystyle{spmpsci}      
\bibliography{main}   

\begin{thebibliography}{10}
\providecommand{\url}[1]{{#1}}
\providecommand{\urlprefix}{URL }
\expandafter\ifx\csname urlstyle\endcsname\relax
  \providecommand{\doi}[1]{DOI~\discretionary{}{}{}#1}\else
  \providecommand{\doi}{DOI~\discretionary{}{}{}\begingroup
  \urlstyle{rm}\Url}\fi

\bibitem{coinmarketcap}
Cryptocurrency market capitalization.
\newblock https://www.coinmarketcap.com/ (2019).
\newblock Accessed: 2019-01-15

\bibitem{abran2004software}
Abran, A., Moore, J.W., Bourque, P., Dupuis, R., Tripp, L.: Software
  engineering body of knowledge.
\newblock IEEE Computer Society, Angela Burgess  (2004)

\bibitem{ADHAMI2018}
Adhami, S., Giudici, G., Martinazzi, S.: Why do businesses go crypto? an
  empirical analysis of initial coin offerings.
\newblock Journal of Economics and Business  (2018)

\bibitem{arkin2005software}
Arkin, B., Stender, S., McGraw, G.: Software penetration testing.
\newblock IEEE Security \& Privacy \textbf{3}(1), 84--87 (2005)

\bibitem{armstrong1977estimating}
Armstrong, J.S., Overton, T.S.: Estimating nonresponse bias in mail surveys.
\newblock Journal of marketing research pp. 396--402 (1977)

\bibitem{azaria2016medrec}
Azaria, A., Ekblaw, A., Vieira, T., Lippman, A.: Medrec: Using blockchain for
  medical data access and permission management.
\newblock In: Open and Big Data (OBD), International Conference on, pp. 25--30.
  IEEE (2016)

\bibitem{beier2005age}
Beier, M.E., Ackerman, P.L.: Age, ability, and the role of prior knowledge on
  the acquisition of new domain knowledge: promising results in a real-world
  learning environment.
\newblock Psychology and aging \textbf{20}(2), 341 (2005)

\bibitem{benjamini1995controlling}
Benjamini, Y., Hochberg, Y.: Controlling the false discovery rate: a practical
  and powerful approach to multiple testing.
\newblock Journal of the royal statistical society. Series B (Methodological)
  pp. 289--300 (1995)

\bibitem{Bosu-etal:JSS}
Bosu, A., Carver, J., Guadagno, R., Bassett, B., McCallum, D., Hochstein, L.:
  Peer impressions in open source organizations: A survey.
\newblock Journal of Systems and Software \textbf{94}(0), 4 -- 15 (2014)

\bibitem{bosu2014impact}
Bosu, A., Carver, J.C.: Impact of developer reputation on code review outcomes
  in oss projects: An empirical investigation.
\newblock In: Proceedings of the 8th ACM/IEEE international symposium on
  empirical software engineering and measurement, p.~33. ACM (2014)

\bibitem{bosu2017process}
Bosu, A., Carver, J.C., Bird, C., Orbeck, J., Chockley, C.: Process aspects and
  social dynamics of contemporary code review: insights from open source
  development and industrial practice at microsoft.
\newblock IEEE Transactions on Software Engineering \textbf{43}(1), 56--75
  (2017)

\bibitem{bosu-emse}
Bosu, A., Iqbal, A., Shahriyar, R., Chakroborty, P.: Understanding the
  motivations, challenges and needs of blockchain software developers: {A}
  survey.
\newblock CoRR \textbf{abs/1811.04169} (2018).
\newblock \urlprefix\url{http://arxiv.org/abs/1811.04169}

\bibitem{testingbcs}
Brooke, S.: The ins and outs of testing blockchain apps (2018).
\newblock
  \urlprefix\url{https://jaxenter.com/ins-outs-testing-blockchain-apps-146447.html}

\bibitem{campion2013system}
Campion, S., Messinger, D.: System and method for distributed software testing
  (2013).
\newblock US Patent 8,621,434

\bibitem{partha2018}
Chakraborty, P., Shahriyar, R., Iqbal, A., Bosu, A.: Understanding the software
  development practices of blockchain projects: A survey.
\newblock In: Proceedings of the 12th ACM/IEEE International Symposium on
  Empirical Software Engineering and Measurement, ESEM '18, pp. 28:1--28:10
  (2018)

\bibitem{chen2017using}
Chen, C., Qi, Z., Liu, Y., Lei, K.: Using virtualization for blockchain
  testing.
\newblock In: International Conference on Smart Computing and Communication,
  pp. 289--299. Springer (2017)

\bibitem{chess2004static}
Chess, B., McGraw, G.: Static analysis for security.
\newblock IEEE Security \& Privacy \textbf{2}(6), 76--79 (2004)

\bibitem{chuen2015handbook}
Chuen, D.L.K.: Handbook of digital currency: Bitcoin, innovation, financial
  instruments, and big data.
\newblock Academic Press (2015)

\bibitem{clack2016smart}
Clack, C.D., Bakshi, V.A., Braine, L.: Smart contract templates: foundations,
  design landscape and research directions.
\newblock arXiv preprint arXiv:1608.00771  (2016)

\bibitem{clarke1996formal}
Clarke, E.M., Wing, J.M.: Formal methods: State of the art and future
  directions.
\newblock ACM Computing Surveys (CSUR) \textbf{28}(4), 626--643 (1996)

\bibitem{Cohen1960}
Cohen, J.: A coefficient of agreement for nominal scales.
\newblock Educational and Psychological Measurement \textbf{20}(1), 37--46
  (1960)

\bibitem{linus2005}
Dahlander, L., Mckelvey, M.: Who is not developing open source software?
  non-users, users, and developers.
\newblock Economics of Innovation and New Technology \textbf{14}(7), 617--635
  (2005)

\bibitem{DAVID2008364}
David, P.A., Shapiro, J.S.: Community-based production of open-source software:
  What do we know about the developers who participate?
\newblock Information Economics and Policy \textbf{20}(4), 364 -- 398 (2008).
\newblock Empirical Issues in Open Source Software

\bibitem{bitcoin-poitics}
De~Filippi, P., Loveluck, B.: The invisible politics of bitcoin: Governance
  crisis of a decentralized infrastructure.
\newblock Internet Policy Review \textbf{5}(4), 529--546 (2016)

\bibitem{deci1985intrinsic}
Deci, E., Ryan, R.M.: Intrinsic motivation and self-determination in human
  behavior.
\newblock Springer Science \& Business Media (1985)

\bibitem{delmolino2016step}
Delmolino, K., Arnett, M., Kosba, A., Miller, A., Shi, E.: Step by step towards
  creating a safe smart contract: Lessons and insights from a cryptocurrency
  lab.
\newblock In: International Conference on Financial Cryptography and Data
  Security, pp. 79--94. Springer (2016)

\bibitem{Destefanis2018BOSE}
Destefanis, G., Marchesi, M., Ortu, M., Tonelli, R., Bracciali, A., Hierons,
  R.M.: Smart contracts vulnerabilities: a call for blockchain software
  engineering?
\newblock In: Proceedings of the 2018 International Workshop on Blockchain
  Oriented Software Engineering, pp. 19--25. IEEE, Campobasso, Italy (2018)

\bibitem{dev2014bitcoin}
Dev, J.A.: Bitcoin mining acceleration and performance quantification.
\newblock In: Electrical and Computer Engineering (CCECE), 2014 IEEE 27th
  Canadian Conference on, pp. 1--6. IEEE (2014)

\bibitem{mainnet}
Documentation, B.D.: Mainnet, bicoin main network.
\newblock \url{https://bitcoin.org/en/glossary/mainnet}.
\newblock Accessed: 2018-01-04

\bibitem{elliott2005obtaining}
Elliott, M.R., Davis, W.W.: Obtaining cancer risk factor prevalence estimates
  in small areas: combining data from two surveys.
\newblock Journal of the Royal Statistical Society: Series C (Applied
  Statistics) \textbf{54}(3), 595--609 (2005)

\bibitem{fanning2016blockchain}
Fanning, K., Centers, D.P.: Blockchain and its coming impact on financial
  services.
\newblock Journal of Corporate Accounting \& Finance \textbf{27}(5), 53--57
  (2016)

\bibitem{ideologyBlog}
Galati, F.: Blockchain as a process: Ideologies and motivations behind the
  technology.
\newblock
  \url{https://medium.com/coinmonks/blockchain-as-a-process-ideologies-and-motivations-behind-the-technology-c25219d87881/}
  (2019)

\bibitem{garfinkel1996public}
Garfinkel, S.L.: Public key cryptography.
\newblock Computer \textbf{29}(6), 101--104 (1996)

\bibitem{coin-dev-index}
Gilbertson, T., Vroegindewey, L.: Larry’s cryptocoin risk (2017).
\newblock \urlprefix\url{https://www.coindevelopmentindex.com/}

\bibitem{hars2001working}
Hars, A., Ou, S.: Working for free? motivations of participating in open source
  projects.
\newblock In: System Sciences, 2001. Proceedings of the 34th Annual Hawaii
  International Conference on, pp. 9--pp. IEEE (2001)

\bibitem{hennessey1998reality}
Hennessey, B.A., Amabile, T.M.: Reality, intrinsic motivation, and creativity.
\newblock American Psychologist \textbf{51}, 1153--1166 (1998)

\bibitem{survey-link}
Herndon, J.B.: Comparing and linking survey data: Considerations for working
  with multiple data sources (2018).
\newblock
  \urlprefix\url{https://www.nidcr.nih.gov/grants-funding/grant-programs/behavioral-social-sciences-research-program/comparing-and-linking-survey-data}

\bibitem{hertel2003motivation}
Hertel, G., Niedner, S., Herrmann, S.: Motivation of software developers in
  open source projects: an internet-based survey of contributors to the linux
  kernel.
\newblock Research policy \textbf{32}(7), 1159--1177 (2003)

\bibitem{hildenbrandt2018kevm}
Hildenbrandt, E., Saxena, M., Rodrigues, N., Zhu, X., Daian, P., Guth, D.,
  Moore, B., Park, D., Zhang, Y., Stefanescu, A., et~al.: Kevm: A complete
  formal semantics of the ethereum virtual machine.
\newblock In: 2018 IEEE 31st Computer Security Foundations Symposium (CSF), pp.
  204--217. IEEE (2018)

\bibitem{hitka2015impact}
Hitka, M., Bal{\'a}{\v{z}}ov{\'a}, {\v{Z}}.: The impact of age, education and
  seniority on motivation of employees.
\newblock Business: Theory and practice \textbf{16}, 113 (2015)

\bibitem{howell2018initial}
Howell, S.T., Niessner, M., Yermack, D.: Initial coin offerings: Financing
  growth with cryptocurrency token sales.
\newblock Tech. rep., National Bureau of Economic Research (2018)

\bibitem{huh2017managing}
Huh, S., Cho, S., Kim, S.: Managing iot devices using blockchain platform.
\newblock In: Advanced Communication Technology (ICACT), 2017 19th
  International Conference on, pp. 464--467. IEEE (2017)

\bibitem{humphrey2007integrating}
Humphrey, S.E., Nahrgang, J.D., Morgeson, F.P.: Integrating motivational,
  social, and contextual work design features: a meta-analytic summary and
  theoretical extension of the work design literature.
\newblock Journal of applied psychology \textbf{92}(5), 1332 (2007)

\bibitem{iansiti2017truth}
Iansiti, M., Lakhani, K.R.: The truth about blockchain.
\newblock Harvard Business Review \textbf{95}(1), 118--127 (2017)

\bibitem{Landis-Koch:1977}
J.~Richard~Landis, G.G.K.: The measurement of observer agreement for
  categorical data.
\newblock Biometrics \textbf{33}(1), 159--174 (1977).
\newblock \urlprefix\url{http://www.jstor.org/stable/2529310}

\bibitem{jacobovitz2016blockchain}
Jacobovitz, O.: Blockchain for identity management (2016)

\bibitem{jakobsson1999proofs}
Jakobsson, M., Juels, A.: Proofs of work and bread pudding protocols.
\newblock In: Secure Information Networks, pp. 258--272. Springer (1999)

\bibitem{jiang2018contractfuzzer}
Jiang, B., Liu, Y., Chan, W.: Contractfuzzer: Fuzzing smart contracts for
  vulnerability detection.
\newblock In: Proceedings of the 33rd ACM/IEEE International Conference on
  Automated Software Engineering, pp. 259--269. ACM (2018)

\bibitem{kononenko2016code}
Kononenko, O., Baysal, O., Godfrey, M.W.: Code review quality: how developers
  see it.
\newblock In: Software Engineering (ICSE), 2016 IEEE/ACM 38th International
  Conference on, pp. 1028--1038. IEEE (2016)

\bibitem{korpela2017digital}
Korpela, K., Hallikas, J., Dahlberg, T.: Digital supply chain transformation
  toward blockchain integration.
\newblock In: Proceedings of the 50th Hawaii international conference on system
  sciences (2017)

\bibitem{Krafft-2018}
Krafft, P.M., Della~Penna, N., Pentland, A.S.: An experimental study of
  cryptocurrency market dynamics.
\newblock In: Proceedings of the 2018 CHI Conference on Human Factors in
  Computing Systems, CHI '18, pp. 605:1--605:13 (2018)

\bibitem{lakhani2003hackers}
Lakhani, K.R., Wolf, R.G.: Why hackers do what they do: Understanding
  motivation and effort in free/open source software projects.
\newblock MIT Sloan Working Paper \textbf{4425-03} (September 2003)

\bibitem{Lee2017}
Lee, A., Carver, J.C., Bosu, A.: Understanding the impressions, motivations,
  and barriers of one time code contributors to floss projects: A survey.
\newblock In: Proceedings of the 39th International Conference on Software
  Engineering, ICSE '17, pp. 187--197. IEEE Press, Piscataway, NJ, USA (2017).
\newblock \doi{10.1109/ICSE.2017.25}.
\newblock \urlprefix\url{https://doi.org/10.1109/ICSE.2017.25}

\bibitem{lohr2000inference}
Lohr, S.L., Rao, J.: Inference from dual frame surveys.
\newblock Journal of the American Statistical Association \textbf{95}(449),
  271--280 (2000)

\bibitem{lu2008rigor}
Lu, C.J., Shulman, S.W.: Rigor and flexibility in computer-based qualitative
  research: Introducing the coding analysis toolkit.
\newblock International Journal of Multiple Research Approaches \textbf{2}(1),
  105--117 (2008)

\bibitem{luu2016making}
Luu, L., Chu, D.H., Olickel, H., Saxena, P., Hobor, A.: Making smart contracts
  smarter.
\newblock In: Proceedings of the 2016 ACM SIGSAC Conference on Computer and
  Communications Security, pp. 254--269. ACM (2016)

\bibitem{mcdonald2007}
Mcdonald, M.P.: {The True Electorate: A Cross-Validation of Voter Registration
  Files and Election Survey Demographics}.
\newblock Public Opinion Quarterly \textbf{71}(4), 588--602 (2007)

\bibitem{meece2006gender}
Meece, J.L., Glienke, B.B., Burg, S.: Gender and motivation.
\newblock Journal of school psychology \textbf{44}(5), 351--373 (2006)

\bibitem{Mockus2002}
Mockus, A., Fielding, R.T., Herbsleb, J.D.: Two case studies of open source
  software development: Apache and mozilla.
\newblock ACM Trans. Softw. Eng. Methodol. \textbf{11}(3), 309--346 (2002)

\bibitem{murphy2014cowboys}
Murphy-Hill, E., Zimmermann, T., Nagappan, N.: Cowboys, ankle sprains, and
  keepers of quality: How is video game development different from software
  development?
\newblock In: Proceedings of the 36th International Conference on Software
  Engineering, pp. 1--11. ACM (2014)

\bibitem{nakamoto2008bitcoin}
Nakamoto, S.: Bitcoin: A peer-to-peer electronic cash system (2008)

\bibitem{narayanan2016bitcoin}
Narayanan, A., Bonneau, J., Felten, E., Miller, A., Goldfeder, S.: Bitcoin and
  Cryptocurrency Technologies: A Comprehensive Introduction.
\newblock Princeton University Press (2016)

\bibitem{nikolic2018finding}
Nikolic, I., Kolluri, A., Sergey, I., Saxena, P., Hobor, A.: Finding the
  greedy, prodigal, and suicidal contracts at scale.
\newblock arXiv preprint arXiv:1802.06038  (2018)

\bibitem{so-survey}
Overflow, S.: Stack overflow annual developer survey (2017).
\newblock \urlprefix\url{https://insights.stackoverflow.com/survey/}

\bibitem{santiagopalladino2017}
Palladino, S.: The parity wallet hack explained (2017).
\newblock
  \urlprefix\url{https://blog.zeppelin.solutions/on-the-parity-wallet-multisig-hack-405a8c12e8f7}

\bibitem{park2018formal}
Park, D., Zhang, Y., Saxena, M., Daian, P., Ro{\c{s}}u, G.: A formal
  verification tool for ethereum vm bytecode.
\newblock In: Proceedings of the 2018 26th ACM Joint Meeting on European
  Software Engineering Conference and Symposium on the Foundations of Software
  Engineering, pp. 912--915. ACM (2018)

\bibitem{peters2016understanding}
Peters, G.W., Panayi, E.: Understanding modern banking ledgers through
  blockchain technologies: Future of transaction processing and smart contracts
  on the internet of money.
\newblock In: Banking Beyond Banks and Money, pp. 239--278. Springer (2016)

\bibitem{porru2017blockchain}
Porru, S., Pinna, A., Marchesi, M., Tonelli, R.: Blockchain-oriented software
  engineering: challenges and new directions.
\newblock In: Proceedings of the 39th International Conference on Software
  Engineering Companion, pp. 169--171. IEEE Press, Buenos Aires, Argentina
  (2017)

\bibitem{posel2006demographics}
Posel, D., Devey, R.: The demographics of fathers in south africa: an analysis
  of survey data, 1993--2002.
\newblock Baba: men and fatherhood in South Africa pp. 38--52 (2006)

\bibitem{raghunathan2007combining}
Raghunathan, T.E., Xie, D., Schenker, N., Parsons, V.L., Davis, W.W., Dodd,
  K.W., Feuer, E.J.: Combining information from two surveys to estimate
  county-level prevalence rates of cancer risk factors and screening.
\newblock Journal of the American Statistical Association \textbf{102}(478),
  474--486 (2007)

\bibitem{Reigers2018ideology}
Reijers, W., Coeckelbergh, M.: The blockchain as a narrative technology:
  Investigating the social ontology and normative configurations of
  cryptocurrencies.
\newblock Philosophy {\&} Technology \textbf{31}(1), 103--130 (2018)

\bibitem{roberts2006understanding}
Roberts, J.A., Hann, I.H., Slaughter, S.A.: Understanding the motivations,
  participation, and performance of open source software developers: A
  longitudinal study of the apache projects.
\newblock Management science \textbf{52}(7), 984--999 (2006)

\bibitem{rosenthal1996qualitative}
Rosenthal, J.A.: Qualitative descriptors of strength of association and effect
  size.
\newblock Journal of social service Research \textbf{21}(4), 37--59 (1996)

\bibitem{rosenthal1994parametric}
Rosenthal, R.: Parametric measures of effect size.
\newblock The handbook of research synthesis pp. 231--244 (1994)

\bibitem{Ross:2010}
Ross, J., Irani, L., Silberman, M.S., Zaldivar, A., Tomlinson, B.: Who are the
  crowdworkers?: Shifting demographics in mechanical turk.
\newblock In: CHI '10 Extended Abstracts on Human Factors in Computing Systems,
  CHI EA '10, pp. 2863--2872 (2010)

\bibitem{sadana2002comparative}
Sadana, R., Mathers, C.D., Lopez, A.D., Murray, C.J., Iburg, K.: Comparative
  analyses of more than 50 household surveys on health status.
\newblock Summary measures of population health: concepts, ethics, measurement
  and applications. Geneva: World Health Organization pp. 369--386 (2002)

\bibitem{schenker2016combining}
Schenker, N., Gentleman, J.F., Rose, D., Hing, E., Shimizu, I.M.: Combining
  estimates from complementary surveys: a case study using prevalence estimates
  from national health surveys of households and nursing homes.
\newblock Public Health Reports  (2016)

\bibitem{shapiro1965}
Shapiro, S.S., Wilk, M.B.: An analysis of variance test for normality (complete
  samples).
\newblock Biometrika \textbf{52}(3/4), 591--611 (1965)

\bibitem{simmons1979symmetric}
Simmons, G.J.: Symmetric and asymmetric encryption.
\newblock ACM Computing Surveys (CSUR) \textbf{11}(4), 305--330 (1979)

\bibitem{snow2013qualtrics}
Snow, J., Mann, M.: Qualtrics survey software: handbook for research
  professionals.
\newblock Qualtrics Labs, Inc  (2013)

\bibitem{spss}
SPSS, I.: Spss statistical software.
\newblock IBM Corporation \textbf{24} (2017)

\bibitem{stallman2003free}
Stallman, R.: Free software foundation (fsf)  (2003)

\bibitem{stewart2006impact}
Stewart, K.J., Gosain, S.: The impact of ideology on effectiveness in open
  source software development teams.
\newblock Mis Quarterly pp. 291--314 (2006)

\bibitem{stoykov2017vibes}
Stoykov, L., Zhang, K., Jacobsen, H.A.: Vibes: fast blockchain simulations for
  large-scale peer-to-peer networks.
\newblock In: Proceedings of the 18th ACM/IFIP/USENIX Middleware Conference:
  Posters and Demos, pp. 19--20. ACM (2017)

\bibitem{suiche2017porosity}
Suiche, M.: Porosity: A decompiler for blockchain-based smart contracts
  bytecode.
\newblock DEF CON \textbf{25} (2017)

\bibitem{swan2015blockchain}
Swan, M.: Blockchain: Blueprint for a new economy.
\newblock " O'Reilly Media, Inc." (2015)

\bibitem{Parity}
Technologies, P.: A postmortem on the parity multi-sig library self-destruct.
\newblock
  \url{https://paritytech.io/a-postmortem-on-the-parity-multi-sig-library-self-destruct/}
  (2017)

\bibitem{securityBestPractice}
technology, S.: Blockchain software security best practices.
\newblock
  \url{https://www.synopsys.com/blogs/software-security/blockchain-software-security-best-practices/}
  (2018)

\bibitem{underwood2016blockchain}
Underwood, S.: Blockchain beyond bitcoin.
\newblock Communications of the ACM \textbf{59}(11), 15--17 (2016)

\bibitem{von2012carrots}
Von~Krogh, G., Haefliger, S., Spaeth, S., Wallin, M.W.: Carrots and rainbows:
  Motivation and social practice in open source software development.
\newblock MIS quarterly pp. 649--676 (2012)

\bibitem{wan2017bug}
Wan, Z., Lo, D., Xia, X., Cai, L.: Bug characteristics in blockchain systems: a
  large-scale empirical study.
\newblock In: Mining Software Repositories (MSR), 2017 IEEE/ACM 14th
  International Conference on, pp. 413--424. IEEE (2017)

\bibitem{west2006challenges}
West, J., Gallagher, S.: Challenges of open innovation: the paradox of firm
  investment in open-source software.
\newblock R\&d Management \textbf{36}(3), 319--331 (2006)

\bibitem{hardfork}
Wiki, B.: Hardfork.
\newblock \url{https://bitcoin.org/en/glossary/hard-fork}.
\newblock Accessed: 2018-01-04

\bibitem{testnet}
Wiki, B.: Testnet.
\newblock \url{https://en.bitcoin.it/wiki/Testnet}.
\newblock Accessed: 2018-01-04

\bibitem{yamane1973statistics}
Yamane, T.: Statistics: An introductory analysis  (1973)

\bibitem{ye2003toward}
Ye, Y., Kishida, K.: Toward an understanding of the motivation open source
  software developers.
\newblock In: Proceedings of the 25th international conference on software
  engineering, pp. 419--429. IEEE Computer Society (2003)

\bibitem{zheng2017overview}
Zheng, Z., Xie, S., Dai, H., Chen, X., Wang, H.: An overview of blockchain
  technology: Architecture, consensus, and future trends.
\newblock In: Big Data (BigData Congress), 2017 IEEE International Congress on,
  pp. 557--564. IEEE (2017)

\bibitem{zheng2016blockchain}
Zheng, Z., Xie, S., Dai, H.N., Wang, H.: Blockchain challenges and
  opportunities: A survey.
\newblock Work Pap.--2016  (2016)

\end{thebibliography}
\nocite{*} \newpage
\appendix


\begin{table*}
	\centering
    \caption{Codes that emerged from our open-coding of the four survey questions and the categories that we assigned each code to.}
	\label{table:codes}
	\resizebox{\linewidth}{!} {
\begin{tabular}{|p{.33\textwidth}|p{.5\textwidth}|p{.3\textwidth}|}
\hline
\textbf{Question}&	\textbf{Codes}&	\textbf{Assigned category }\\ \hline \hline
\multirow{ 8}{0.95\linewidth}{
Q10: What are your motivations to contribute to your primary project?} 	&Ideology &	Ideology\\ \hhline{~--}
	&Financial gains&	\multirow{ 2}{\linewidth}{External rewards}\\ \hhline{~-~}
	&Job, profession	& \\ \hhline{~--}
& 	Hobby, Fun& \multirow{ 2}{\linewidth}{	Intrinsic motivation} \\ \hhline{~-~}
	&Passion, Self interest&	\\ \hhline{~--}
	&Technical attraction	&Technical attraction\\ \hhline{~--}
	&Learning, professional development	&Learning\\ \hhline{~--}
	&Community recognition&	Community Recognition\\ \hline
	\hline
\multirow{14}{0.95\linewidth}{Q11: Based on your experiences, what are the primary differences between blockchain and non-blockchain software development ?} &High emphasis on security reliability&	\multirow{3}{\linewidth}{Security/ reliability}\\ \hhline{~-~}
	&Irreversible data	&\\ \hhline{~-~}
	&Costly defects	&\\ \hhline{~--}		
	
&	Networking knowledge &	\multirow{ 4}{\linewidth}{Domain characteristics}\\ \hhline{~-~}
	&Cryptography knowledge	&\\ \hhline{~-~}
	&Distributed environment&	\\ \hhline{~-~}
	&High pace development	&\\ \hhline{~--}
	&Lack of tools&	\multirow{3}{\linewidth}{Immature ecosystem}\\ \hhline{~-~}
	&New technology /framework	&\\ \hhline{~-~}
	&Lack of experienced developer&	\\ \hhline{~--}
	&Backward compatibility &	\multirow{ 2}{\linewidth}{Maintainability}\\ \hhline{~-~}
	&Upgrade difficulty	&\\ \hhline{~--}
		&No difference&	No difference \\ \hhline{~--}
	& Others &Others \\ \hline
	\hline
\multirow{14}{0.95\linewidth}{Q13: What are the most challenging aspects of blockchain software development?} 	&Cost of defects	&\multirow{ 6}{\linewidth}{Uncommon characteristics of BCS}\\ \hhline{~-~}
	&Steep learning curve&	\\ \hhline{~-~}
	&Technological complexity&	\\ \hhline{~-~}
	&Maintainability	&\\ \hhline{~-~}
	&High pace development	&\\ \hhline{~-~}
	&Scalability&	\\ \hhline{~--}
&	Testing&	\multirow{3}{\linewidth}{Development challenges}\\ \hhline{~-~}
	&Security	&\\ \hhline{~-~}
	&Code reviews&	\\ \hhline{~--}
	&Collaboration difficulties	&\multirow{3}{\linewidth}{Non-technical challenges}\\ \hhline{~-~}
	&Governance, politics	&\\ \hhline{~-~}
	&Ethical aspects	&\\ \hhline{~--}
		&Lack of tools	&\multirow{ 2}{\linewidth}{Lack of supporting materials}\\ \hhline{~-~}
	&Lack of documentation	& \\ \hline \hline
		
\multirow{14}{0.95\linewidth}{Q14: Please describe the type of tools that you currently do not have, but if implemented, can greatly help your blockchain software development activities.}
	&Simulator&	\multirow{ 5}{\linewidth}{Testing supports}\\ \hhline{~-~}
	&Security testing support	&\\ \hhline{~-~}
	&Easily deployable testnet	&\\ \hhline{~-~}
	&Formal spec verifier	&\\ \hhline{~-~}
	&Static analysis tools	&\\ \hhline{~--}
&	Development environment /IDE&	\multirow{3}{\linewidth}{Development supports}\\ \hhline{~-~}
	&Stable development tools	&\\ \hhline{~-~}
	&Easy to write formal specification	&\\ \hhline{~--}

	&IDE for smart contracts	& \multirow{3}{\linewidth}{Smart-contract support}\\ \hhline{~-~}
	&Debugger for smart contracts	&\\ \hhline{~-~}
	&EVM decompiler	&\\ \hhline{~--}
	&Blockchain specific design notations&	\multirow{3}{\linewidth}{Others}\\ \hhline{~-~}
	&Container	&\\ \hhline{~-~}
	&Others	& \\ \hline

\end{tabular}
}
\end{table*}

\end{document}